\documentclass[twocolumn,bibnotes,showpacs,amsmath,amssymb,aps,prl]{revtex4-2}

\usepackage{graphicx}
\usepackage{amsmath}
\usepackage{amssymb}
\usepackage{color}
\usepackage{comment}
\usepackage[normalem]{ulem}
\usepackage{bm}

\usepackage[colorlinks=true,linkcolor=blue,citecolor=blue]{hyperref}

\begin{document}
\title{Two-step growth of high-quality single crystals of the Kitaev magnet $\alpha$-RuCl$_{3}$}
\author{R.~Namba$^{1,{\S}}$}
\author{K.~Imamura$^{1,{\S}}$}
\author{R.~Ishioka$^1$}
\author{K.~Ishihara$^1$}
\author{T.~Miyamoto$^1$}
\author{H.~Okamoto$^1$}
\author{Y.~Shimizu$^2$}
\author{Y.~Saito$^3$}
\author{Y.~Agarmani$^3$}
\author{M.~Lang$^3$}
\author{H.~Murayama$^4$}
\author{Y.~Xing$^5$}
\author{S.~Suetsugu$^6$}
\author{Y.~Kasahara$^6$}
\author{Y.~Matsuda$^6$}
\author{K.~Hashimoto$^1$}\email{k.hashimoto@edu.k.u-tokyo.ac.jp}
\author{T.~Shibauchi$^1$}\email{shibauchi@k.u-tokyo.ac.jp}
\affiliation{
$^1$Department of Advanced Materials Science, University of Tokyo, Kashiwa, Chiba 277-8561, Japan\\
$^2$Department of Physics, Nagoya University, Furo-cho, Chikusa-ku, Nagoya 464-8602, Japan\\
$^3$Institute of Physics, Goethe University Frankfurt, 60438 Frankfurt (M), Germany\\
$^4$RIKEN Center for Emergent Matter Science (CEMS), Wako 351-0198, Japan\\
$^5$State Key Laboratory of Heavy Oil Processing, College of New Energy and Materials, China University of Petroleum, Beijing 102249, People's Republic of China\\
$^6$Department of Physics, Kyoto University, Kyoto 606-8502, Japan\\
$^{\S}$\rm These authors contributed equally to this work.
}

\begin{abstract}
The layered honeycomb magnet $\alpha$-RuCl$_3$ is the most promising candidate for a Kitaev quantum spin liquid (KQSL) that can host charge-neutral Majorana fermions. Recent studies have shown significant sample dependence of thermal transport properties, which are a key probe of Majorana quasiparticles in the KQSL state, highlighting the importance of preparing high-quality single crystals of $\alpha$-RuCl$_3$. Here, we present a relatively simple and reliable method to grow high-quality single crystals of $\alpha$-RuCl$_3$. We use a two-step crystal growth method consisting of a purification process by chemical vapor transport (CVT) and a main crystal growth process by sublimation. The obtained crystals exhibit a distinct first-order structural phase transition from the monoclinic ($C2/m$) to the rhombohedral ($R\bar{3}$) structure at $\sim150$\,K, which is confirmed by the nuclear quadrupole resonance spectra with much sharper widths than previously reported. The Raman spectra show the absence of defect-induced modes, supporting the good crystallinity of our samples. The jumps in the thermal expansion coefficient and specific heat at the antiferromagnetic (AFM) transition at 7.6-7.7\,K are larger and sharper than those of previous samples grown by the CVT and Bridgman methods and do not show any additional AFM transitions at 10-14\,K due to stacking faults. The longitudinal thermal conductivity in the AFM phase is significantly larger than previously reported, indicating a very long mean free path of heat carriers. All the results indicate that our single crystals are of superior quality with good crystallinity and few stacking faults, which provides a suitable platform for studying the Kitaev physics.
\end{abstract}

\maketitle
\section{Introduction}
Quantum spin liquids (QSLs) are exotic quantum states in which spins are strongly correlated with each other but do not order down to absolute zero due to strong frustrations~\cite{Balents2010}. Following the proposal of a QSL state on a triangular lattice by Anderson~\cite{Anderson1973}, various approaches have been proposed to stabilize QSL states. Among them, the Kitaev model on a two-dimensional (2D) honeycomb lattice has garnered considerable attention because it can host an exactly solvable QSL state~\cite{Kitaev2006}. In this model, bond-dependent Ising interactions introduce exchange frustrations, giving rise to a Kitaev quantum spin liquid (KQSL) state, where the spin degrees of freedom fractionalize into itinerant Majorana fermions and localized $Z_2$ fluxes (visons)~\cite{Motome2020}. Intriguingly, these quasiparticles form composite particles under magnetic fields, which are expected to be non-Abelian anyons that are indispensable for fault-tolerant topological quantum computations~\cite{Kitaev2006}.

After the proposal of the Kitaev model, Jackeli and Khaliullin proposed an innovative mechanism to realize the Kitaev interactions in real materials~\cite{Jackeli2009}. Since then, many candidate materials have been proposed~\cite{Takagi2019}. Among them, the spin-orbit-assisted Mott insulator $\alpha$-RuCl$_{3}$ with a layered honeycomb lattice~\cite{Plumb2014} has been most actively studied, in which Ru$^{3+}$ ions are surrounded by octahedra of Cl$^-$ ions. Since the adjacent octahedra share the edge of the Cl-Cl bond, the Heisenberg interactions through the two Ru-Cl-Ru bonds are canceled except for the Ising interactions perpendicular to the edge directions, providing a promising platform for realizing the Kitaev model. $\alpha$-RuCl$_{3}$ exhibits an antiferromagnetic (AFM) order below $T_{\mathrm N} \sim$7$\,$K due to non-Kitaev interactions, such as Heisenberg and off-diagonal interactions~\cite{Kubota2015,Majumder2015,Johnson2015,Winter2016}, but various experiments above $T_{\mathrm N}$, including neutron scattering~\cite{Banerjee2016proximate,Banerjee2017,Do2017}, Raman spectroscopy~\cite{Sandilands2015,Nasu2016}, and specific heat~\cite{Do2017,Widmann2019}, suggest the spin fractionalization expected in the KQSL state. More importantly, the AFM phase can be suppressed by in-plane magnetic fields above $\sim$8$\,$T~\cite{Yadav2016,Wolter2017,Banerjee2018}, leading to a field-induced spin-disordered state. In this region, the half-integer quantized thermal Hall effect has been observed~\cite{Kasahara2018,Yokoi2021,Yamashita2020,Bruin2022,Kasahara2022}, supporting the existence of chiral edge currents of Majorana fermions at the sample edges. Moreover, recent specific heat measurements under in-plane magnetic fields demonstrate the presence of a Majorana gap with characteristic angular dependence expected in the Kitaev model~\cite{Tanaka2022,Imamura2023}. The sign change of the thermal Hall effect is observed at the field angle where the Majorana gap closes~\cite{Imamura2023}, providing compelling evidence for a one-to-one correspondence between the bulk and edge properties expected for the topological state in the KQSL under magnetic fields.

However, it has been pointed out that the experimental value of the thermal Hall conductivity $\kappa_{xy}$ strongly depends on the sample quality and crystal growth methods. For instance, it has been reported that samples synthesized by the Bridgman method exhibit the half-integer quantized thermal Hall effect~\cite{Kasahara2018,Yokoi2021,Yamashita2020,Bruin2022,Kasahara2022}, whereas samples synthesized by the vapor transport method do not exhibit a perfect half-integer quantization value, leading to a proposal of different interpretation based on a thermal Hall effect due to topological magnons~\cite{czajka2023}. In addition, the observation of scaling in $\kappa_{xx}$ and $\kappa_{xy}$ points to the possibility of the phonon thermal Hall effect~\cite{lefranccois2022}, but such scaling has not been observed in other groups~\cite{Kasahara2018,Bruin2022,czajka2023}. Moreover, even samples grown by the same method show strong sample dependence; for instance, in Bridgman samples, the half-integer quantized thermal Hall effect tends to be observed in cleaner samples with fewer stacking faults and higher longitudinal thermal conductivity $\kappa_{xx}$~\cite{Kasahara2022}.

In $\alpha$-RuCl$_{3}$, apart from the half-integer quantum thermal Hall effect in $\kappa_{xy}$, it has been reported that the longitudinal thermal conductivity $\kappa_{xx}$ oscillates periodically with respect to the inverse of the magnetic field~\cite{czajka2021}. However, the interpretation remains highly controversial. Some claim that these are quantum oscillations resulting from an exotic QSL state~\cite{czajka2021}, which may be associated with the KQSL state~\cite{Zhang2023}, while others argue that they are due to additional AFM transitions caused by stacking faults~\cite{Bruuin_oscillation}. Therefore, it is highly desirable to investigate the effects of sample-quality dependence using high-quality single crystals to examine whether or not the KQSL state is realized in $\alpha$-RuCl$_{3}$. 

There have been several reports on the synthesis and characterization of $\alpha$-RuCl$_{3}$ single crystals, and it is well-known that crystal growth methods involving a sublimation process, such as the self-selective vapor phase growth (SSVG) method~\cite{Yan_SSVG_2023,Zhang2024}, can yield high-quality single crystals. In particular, it has been reported that high-quality single crystals can be obtained by the sublimation method using preannealed $\alpha$-RuCl$_{3}$ powder~\cite{Wolf2022}. Here, we report on the synthesis of high-quality single crystals of $\alpha$-RuCl$_{3}$ by a two-step crystal growth method using the CVT and sublimation processes. We carefully characterized the crystal structure of our samples and investigated the effects of sample quality on the physical properties of $\alpha$-RuCl$_{3}$ through magnetic susceptibility, nuclear quadrupole resonance (NQR), Raman spectroscopy, thermal expansion, specific heat, and longitudinal thermal conductivity measurements. A clear structural phase transition from the monoclinic ($C2/m$) to the rhombohedral ($R\bar{3}$) structure was observed at 140-160$\,$K. The jumps in the thermal expansion coefficient and specific heat at the AFM transition temperature $T_{\mathrm{N}}$ are significantly higher and sharper than those of previous Bridgman~\cite{Tanaka2022,Imamura2023} and CVT~\cite{Majumder2015,Mi2015} samples. Moreover, the longitudinal thermal conductivity $\kappa_{xx}$ in the AFM phase is the highest among the samples reported so far. All the results consistently indicate that our single crystals are of high quality with good crystallinity and few stacking faults.

\section{Experiment}
\begin{figure}[b]
    \includegraphics[width=0.95\linewidth]{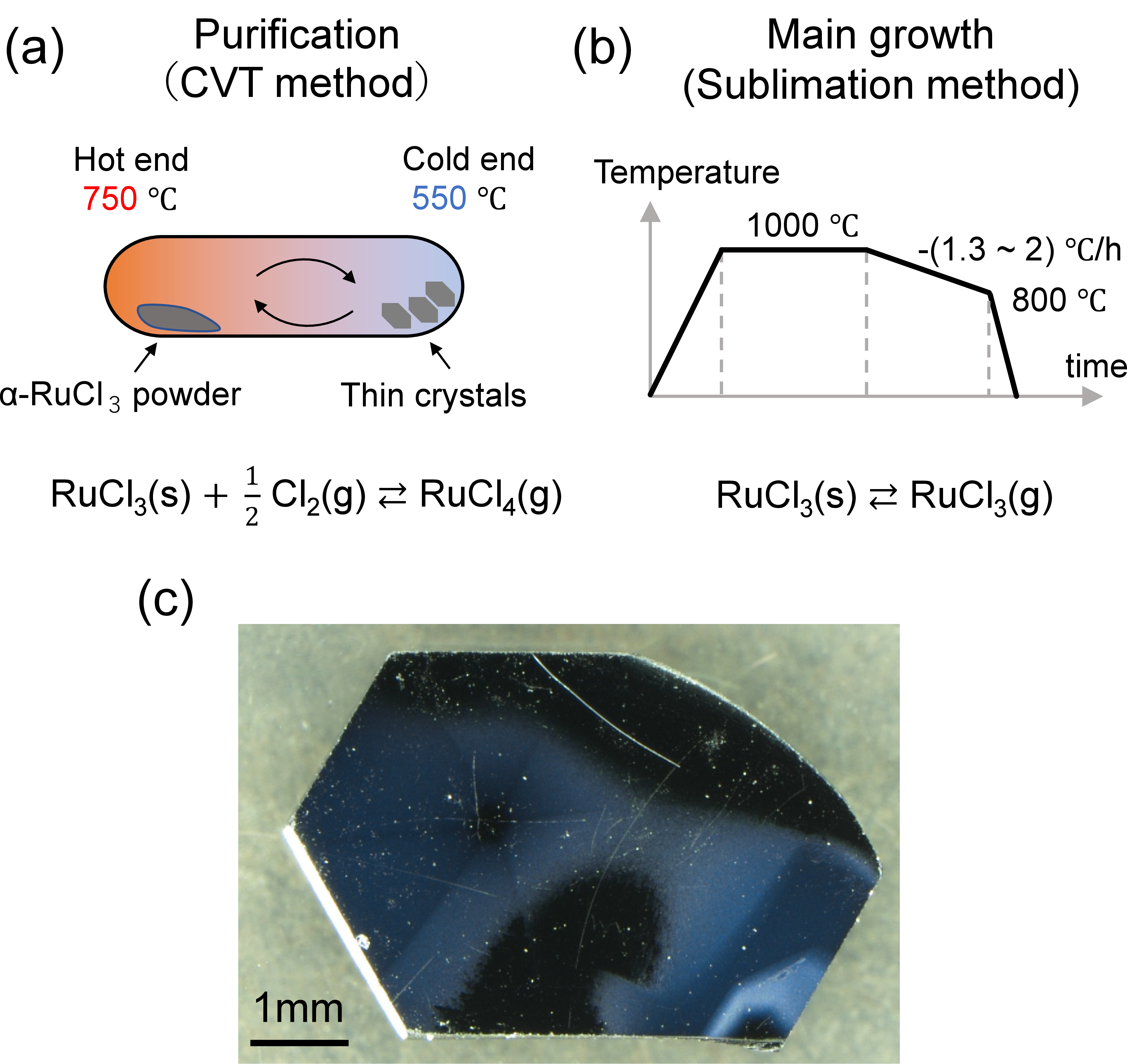}
    \caption{(a) Purification process by the CVT method. The starting powder of $\alpha$-RuCl$_{3}$ are transported as RuCl$_4$ gas from the hot end to the cold end. Tiny single crystals are grown at the cold end, while impurities such as oxides are retained at the hot end. (b) Main crystal growth process by the sublimation method. A typical temperature sequence is shown. (c) Photograph of a typical $\alpha$-RuCl$_{3}$ single crystal obtained by the two-step sublimation method.}
    \label{Fig1}
\end{figure}
In this study, single crystals of $\alpha$-RuCl$_{3}$ were synthesized by a two-step process. First, the starting powder of polycrystalline $\alpha$-RuCl$_{3}$ (Furuya Metal) was carefully purified and crystallized by the CVT method. In this process, no external transport agent is used because the $\alpha$-RuCl$_{3}$ powder dissociates, producing chlorine gas, which acts as a transport agent. It is known that single crystals of $\alpha$-RuCl$_{3}$ synthesized by the CVT method, where crystal growth occurs at relatively lower temperatures compared to other methods, tend to contain stacking faults and have lower crystallinity~\cite{Johnson2015,May_2020}. However, this process is useful for removing impurities, such as oxides, from the starting powder~\cite{Cao2016}. In the first process, the starting powder with a weight of 0.26\,g is placed at the hot end of a quartz tube (typically, 15\,mm in out diameter, 26\,cm long, and 1\,mm thick) and transported to the cold end, as illustrated in Fig.$\,$\ref{Fig1}(a). The temperatures at the hot and cold ends were set at 750$\,$${}^\circ$C and 550$\,$${}^\circ$C, respectively, which are typical values for the previous CVT method applied to $\alpha$-RuCl$_{3}$~\cite{Kim2022}. This process yielded tiny plate-like single crystals at the cold end. The obtained CVT crystals then served as starting materials for the subsequent sublimation growth process, which was conducted without a temperature gradient. In this main growth process, CVT crystals were enclosed inside a quartz tube (typically, 15\,mm in out diameter, 8\,cm long, and 1\,mm thick). The total volume of the CVT crystals (typically, 45\,mg) was set so that the gas pressure inside the quartz tube reached 2-3\,atm at 1000\,${}^\circ$C. Single crystals enclosed in a quartz tube were held at 1000$\,$${}^\circ$C for 24\,h and then gradually cooled to 800$\,$${}^\circ$C at a rate of $-(1.3\sim 2)\,^\circ$C/h (Fig.$\,$\ref{Fig1}(b)). Figure$\,$\ref{Fig1}(c) displays a typical single crystal obtained by this two-step sublimation method. Typically, all the tiny starting single crystals crystallize into one single crystal during the sublimation process, resulting in a millimeter-sized large crystal with black, shiny surfaces. 
In the main crystal growth process by the sublimation method, we placed six quartz tubes in a furnace in one process. Finally, we obtained single crystals in almost all the quartz tubes (about 5 out of 6).

The crystal structure at room temperature was evaluated by X-ray diffraction measurements (XRD) (RIGAKU MicroMax-007 HF). The magnetic susceptibility was measured on a superconducting quantum interference device magnetometer (Quantum Design MPMS XL). The thermal expansion coefficient was measured with a homemade capacitive dilatometer~\cite{pott1983apparatus,Wolf2022}. Our apparatus enables the detection of length changes $\Delta L\geq 10^{-2}$$\mbox{\AA}$, where $L$ is the length of the sample. NQR measurements were performed by the conventional spin-echo method~\cite{Nagai_NQRNMR_2020}. The Raman spectroscopy was measured using a confocal microscopic system with a linear polarized laser (Horiba LabRAM HR Evolution). The excitation wavelength was 532\,nm. Specific heat measurements were carried out by the long-relaxation technique~\cite{Tanaka2022}. The longitudinal thermal conductivity was measured by the standard steady-state method~\cite{Yokoi2021}.

\section{Results and Discussion}
\begin{figure}[b]
    \includegraphics[width=0.85\linewidth]{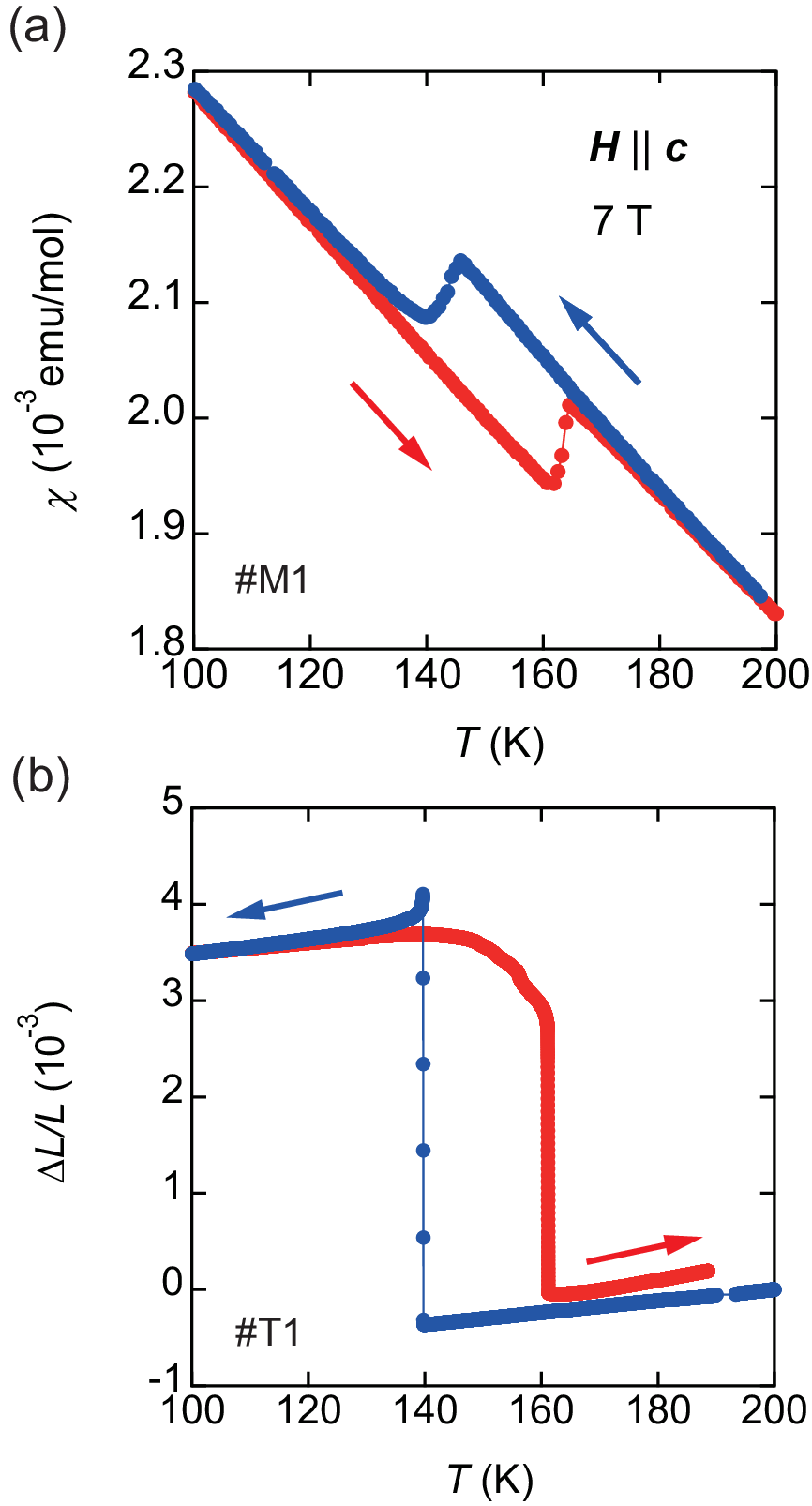}
    \caption{(a) Temperature dependence of magnetic susceptibility of $\alpha$-RuCl$_{3}$ under a magnetic field of 7\,T measured during cooling (blue) and heating (red) processes around at $T_{\mathrm s}$. (b) Temperature dependence of relative length change of $\alpha$-RuCl$_{3}$ measured along the $a$-axis around at $T_{\mathrm s}$ during cooling (blue) and heating (red) processes.}
    \label{Fig2}
\end{figure}

\begin{figure*}[t]
    \includegraphics[width=1\linewidth]{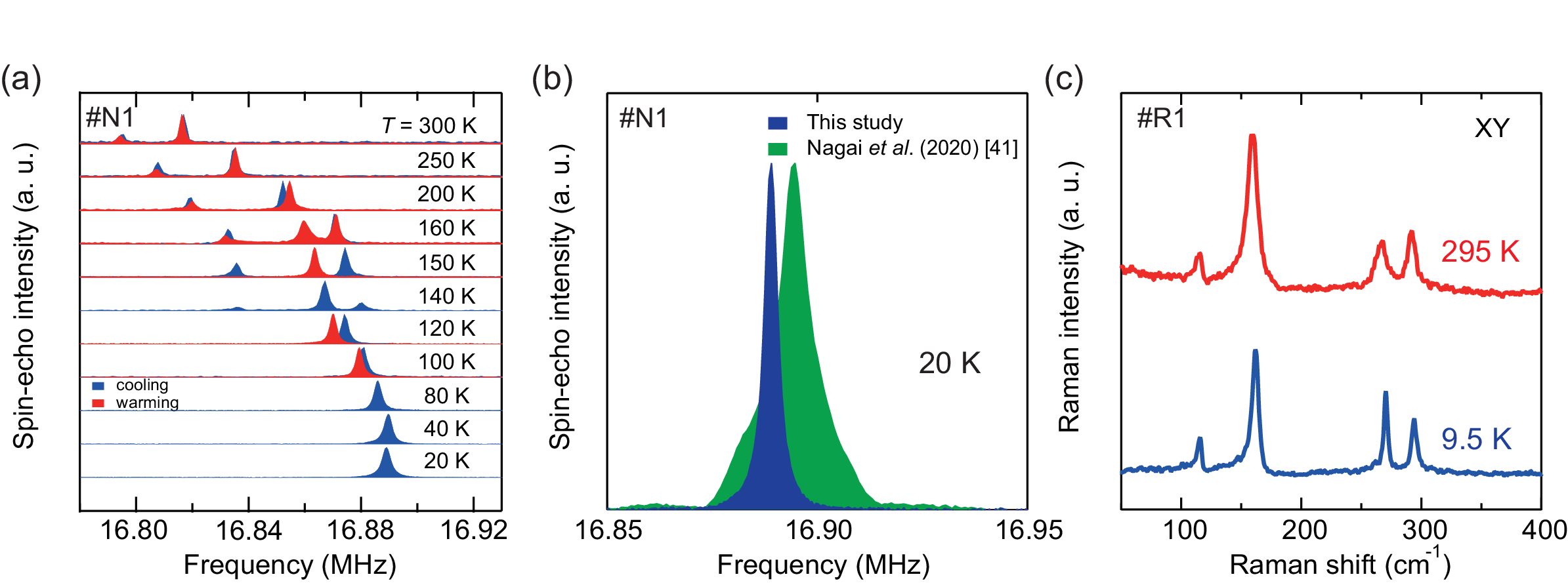}
    \caption{(a) Temperature dependence of $^{35}$Cl NQR spectra of $\alpha$-RuCl$_{3}$ measured during cooling (blue) and heating (red) processes. (b) Enlarged view of the $^{35}$Cl NQR spectra at 20\,K for the sample measured in this study (blue) and for the sample grown by the single-step sublimation method (green)~\cite{Nagai_NQRNMR_2020}. (c) Raman spectra of $\alpha$-RuCl$_{3}$ measured at 295\,K (red) and 9.5\,K (blue) in the $z(xy)\bar{z}$ configuration with in-plane linear polarization (XY geometry). Here, $x$ and $y$ correspond to the $a$- and $b$-axes in the monoclinic $C2/m$ structure, respectively, and $z$ denotes the direction of the incident light normal to the $ab$ plane.}
    \label{Fig3}
\end{figure*}

It is well-established that $\alpha$-RuCl$_{3}$ undergoes a structural phase transition at $T_{\mathrm{s}}$\,=\,140-160\,K~\cite{Kubota2015}. However, the crystal structures both above and below $T_{\mathrm s}$ are still a subject of debate. The space group of the crystal structure is considered to be $C2/m$ at room temperature and $R\Bar{3}$ below $T_{\mathrm s}$~\cite{Kim2024}. However, several previous studies have argued that the room-temperature structure is $P3_112$, and the low-temperature structure is either $C2/m$ or $P3_112$~\cite{Bruuin_oscillation,Cao2016,Johnson2015}. Generally, Bragg peaks in XRD measurements are influenced by stacking faults and multidomains in crystals. $\alpha$-RuCl$_{3}$ is susceptible to stacking faults because the 2D honeycomb planes are weakly stacked by the interlayer van der Waals force~\cite{Cao2016}. Additionally, in the monoclinic $C2/m$ crystal structure, multidomains can form, which may become a source of the partial cancellation of the thermal Hall signal~\cite{Kurumaji2023}. Therefore, the contradiction of the crystal structure in $\alpha$-RuCl$_{3}$ may be attributed to stacking faults and multidomains. For instance, the room-temperature structure with $P3_112$ can be observed due to multidomain structures with $C2/m$~\cite{Johnson2015}, and the low-temperature structure with $C2/m$ or $P3_112$ can be observed when defects in the crystal hinder the structural phase transition, maintaining the room-temperature structure. Thus, in order to address the KQSL state of $\alpha$-RuCl$_{3}$, it is essential to prepare high-quality single crystals with fewer stacking faults and multidomains.

The XRD measurements reveal that the crystal structure of our samples is the monoclinic $C2/m$ structure at room temperature with lattice constants $a=5.9956(12)$\,$\mbox{\AA}$, $b=10.354(17)$\,$\mbox{\AA}$, $c=6.0237(11)$\,$\mbox{\AA}$, and $\beta=108.83(2)$\,$^{\circ}$. Figures\,\ref{Fig2}(a) and (b) depict the temperature dependence of the magnetic susceptibility $\chi(T)$ under a magnetic field of 7$\,$T along the $c$-axis (perpendicular to the 2D honeycomb plane) and the relative length change $\Delta L(T)/L(T_0)$, where $\Delta L(T)\equiv L(T)-L(T_0)$ and $T_0$ is the reference temperature, measured along the $a$-axis (perpendicular to the Ru-Ru bond direction) during cooling and heating, respectively. Both datasets exhibit a distinct first-order structural phase transition with a hysteresis loop within a narrower temperature range of 20\,K compared to the samples synthesized by the vapor transport method\,\cite{He_2018}, indicating the high quality of our samples. As discussed below, crystals of higher quality, characterized by a clear structural phase transition, demonstrate a higher antiferromagnetic transition temperature $T_{\mathrm N}$.

Next, we performed NQR measurements under zero magnetic field from 300\,K to 20\,K to obtain detailed information on the crystal structure. Figure \,\ref{Fig3}(a) displays the NQR spectra of $^{35}$Cl ($I = 3/2$) at around 16.8-16.9\,MHz. The number of peaks in the NQR spectra corresponds to the number of independent sites of Cl, reflecting the crystal structure of the material. Above 160\,K, two peaks with different intensities are observed, consistent with the $C2/m$ crystal structure. Below 120\,K, only one peak is observed, indicating the \textit{R}$\Bar{3}$ crystal structure. This result coincides with that of CrCl$_{3}$~\cite{Morosin_1964}, which has the same structure as $\alpha$-RuCl$_{3}$. In the intermediate temperature range of 140-160\,K, three peaks with different intensities are observed, which are interpreted as the coexistence of the high-temperature $C2/m$ structure and the low-temperature \textit{R}$\Bar{3}$ structure. If the crystal structure is \textit{P}3$_1$12, three peaks should be observed in the NQR spectra~\cite{Morosin_1964}, but neither three peaks above nor below $T_{\mathrm s}$ are observed. Therefore, the possibility of the crystal structure with \textit{P}3$_1$12 in $\alpha$-RuCl$_{3}$ is excluded. This result is consistent with recent XRD results showing that the crystal structure at room temperature is $C2/m$ and a structural phase transition from $C2/m$ to \textit{R}$\Bar{3}$ occurs at $T_{\rm s}$~\cite{Kim2024}.
Figure$\,$\ref{Fig3}(b) compares the NQR spectra of our sample with those of a sample synthesized by the one-step sublimation method without purification by the CVT process~\cite{Nagai_NQRNMR_2020}. The linewidth of our sample is less than half of that of the previous sample. This also confirms that our sample is of high quality.

Despite the pronounced change at $T_{\mathrm s}$ discussed above, the in-plane Raman spectra in Fig.$\,$\ref{Fig3}(c) exhibit nearly identical phonon peak structures between 295\,K and 9.5\,K. This reflects the layered structure of $\alpha$-RuCl$_{3}$ with weak interlayer van der Waals interactions. Consequently, the in-plane Raman spectra depend only on the structure of the isolated $\alpha$-RuCl$_{3}$ layer. In accordance with a previous study~\cite{Sandilands2015}, four Raman active modes are expected in the XY geometry. Indeed, four Raman modes are observed in our sample, indicating that the interlayer lattice interactions are indeed weak in $\alpha$-RuCl$_{3}$ and do not significantly affect the phonon dynamics expected in the isolated RuCl$_{3}$ layer. In the previous study~\cite{Sandilands2015}, an additional peak at 205\,cm$^{-1}$ was observed alongside the four peaks, interpreted as being activated by defects. Notably, such a peak is absent in our sample, affirming the high quality of our sample.

\begin{figure}[t]
    \includegraphics[width=1\linewidth]{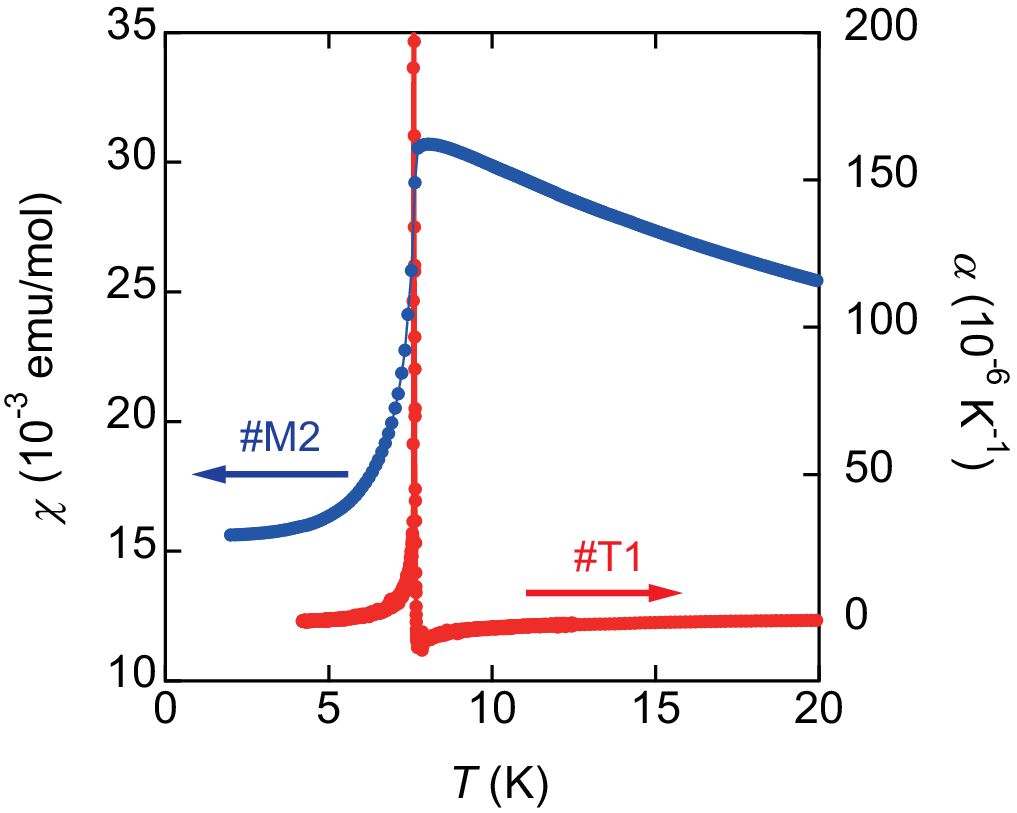}
    \caption{Temperature dependence of magnetic susceptibility $\chi$ under a magnetic field of 0.1\,T perpendicular to the $c$-axis (blue, left axis) and thermal expansion coefficient $\alpha$ (red, right axis) measured along the $a$-axis at around $T_{\mathrm N}$.
    }
    \label{Fig4}
\end{figure}

Figure\,\ref{Fig4} illustrates the temperature dependence of the in-plane magnetic susceptibility $\chi$ and the thermal expansion coefficient $\alpha(\equiv L^{-1}{\rm d}L/{\rm d}T)$ at low temperatures, revealing a distinct AFM transition at $T_{\mathrm N}  = 7.6$\,K. In particular, $\alpha(T)$ exhibits a sharper AFM transition compared to the previous samples synthesized by the vapor transport method~\cite{He_2018}, although the sign of the change in $\alpha$ at $T_{\rm N}$ for our high-quality sample is opposite to that of the previous samples with relatively broader transitions at $T_{\rm N}$ and $T_{\rm s}$. Further studies are needed to clarify this point. In addition to the sharper transition at $T_{\rm N}$, the AFM transition temperature in our samples is among the highest reported so far, indicating the high purity of our samples. $\alpha$-RuCl$_{3}$ has an ABC stacking structure below $T_{\rm{s}}$, but when stacking faults are introduced, it changes to an ABAB stacking structure, resulting in additional AFM orders at around 10-14\,K~\cite{Cao2016}. In our samples, no obvious anomaly is observed around 10-14 K, which indicates minimal stacking faults in our samples.

\begin{figure}[t]
    \includegraphics[width=0.85\linewidth]{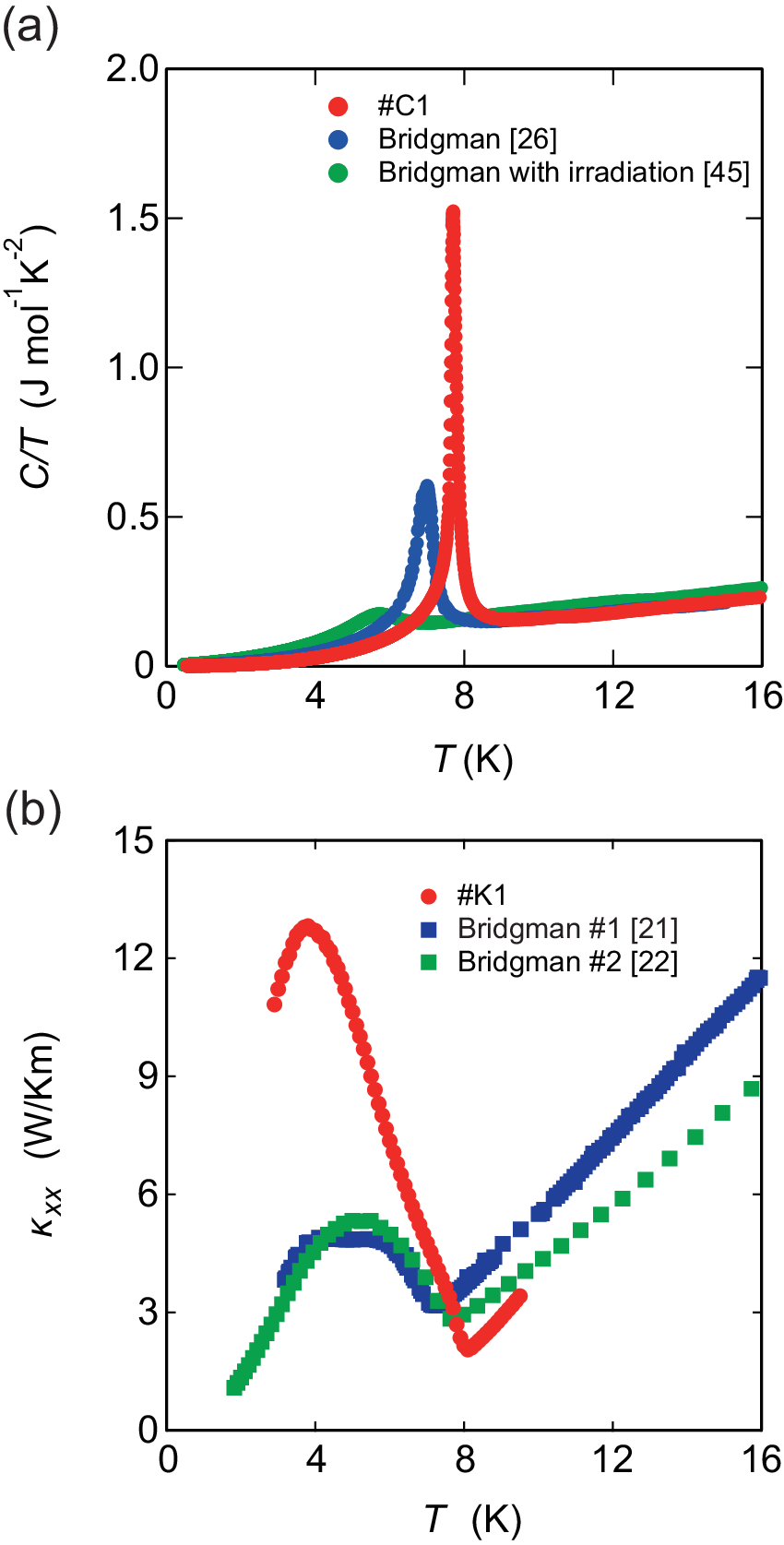}
    \caption{(a) Temperature dependence of specific heat divided by temperature $C/T$ for the two-step sublimation sample (\#C1, red) and Bridgman samples with (green) and without (blue) electron irradiation~\cite{Tanaka2022,imamura2023irradiation}. (b) Temperature dependence of longitudinal thermal conductivity $\kappa_{xx}$ along the $a$-axis for the two-step sublimation sample (\#K1, red) compared with Bridgman samples (blue and green) showing the half-integer quantized thermal Hall effect~\cite{Yokoi2021,Kasahara2018}.}
    \label{Fig5}
\end{figure}

Figure\,\ref{Fig5}(a) shows the specific heat divided by temperature $C/T$. Again, a clear AFM transition at $T_{\mathrm{N}}=7.7$\,K is observed in our sample, which is higher than that of previous Bridgman~\cite{Tanaka2022,Imamura2023} and CVT~\cite{Majumder2015,Mi2015} samples. The peak of the specific heat $C/T$ at $T_{\mathrm N}$ is much sharper and higher than that of the previous Bridgman sample~\cite{Tanaka2022,Imamura2023} and comparable to that of the SSVG sample~\cite{Zhang2024}. 
For comparison, we have plotted in Fig.\,\ref{Fig5}(a) the $C/T$ data for Bridgman samples with and without point defects introduced by electron irradiation~\cite{imamura2023irradiation}. The introduction of defects by electron irradiation leads to a shift in $T_{\mathrm N}$ towards lower temperatures, accompanied by a suppression of the sharpness of the transition at $T_{\mathrm N}$. According to the previous studies~\cite{Zhang2024}, samples with higher $T_{\mathrm N}$ exhibit sharp Bragg peak spots in XRD measurements and a clear structural phase transition at $T_{\mathrm s}$. Moreover, no anomalies due to stacking faults have been observed in the specific heat at 10-14\,K. These results indicate the exceptional crystallinity of our samples.

Finally, we present the results of longitudinal thermal conductivity measurements. Figure\,\ref{Fig5}(b) depicts the temperature dependence of the thermal conductivity $\kappa_{xx}$ along the $a$-axis, together with data for Bridgman samples reported in previous studies that exhibit the half-integer quantized thermal Hall effect~\cite{Yokoi2021,Kasahara2018}. It has been argued for the Bridgman samples that cleaner samples with a large peak of thermal conductivity $\kappa_{xx}^{\mathrm{peak}}$ above 4\,W/Km in the AFM phase tend to exhibit a value close to the half-integer quantization value~\cite{Yokoi2021,Kasahara2022}. In the AFM phase, spin-phonon scattering is suppressed, and the phonon contribution to the thermal conduction becomes enhanced. Since the enhancement is more pronounced for cleaner samples with smaller impurity scattering, the magnitude of $\kappa_{xx}^{\mathrm{peak}}$, which is proportional to the mean free path of the heat carriers, is a measure of crystallinity. The present sample exhibits a much higher $\kappa_{xx}^{\mathrm{peak}}$ than that of the Bridgman samples exhibiting the half-integer quantized thermal Hall effect. The obtained value is more significant than that of clean samples synthesized by the SSVG method~\cite{Zhang2024,Yan_SSVG_2023}, which is the highest value reported so far. In contrast, $\kappa_{xx}$ in the paramagnetic state above 8\,K is smaller than that of samples grown by other crystal growth methods. This can be attributed to stronger spin-phonon scattering in the high-quality crystals in the paramagnetic phase, aligning with the previous report~\cite{Zhang2024}. In Table\,\ref{Table1}, we summarize the results for $T_{\mathrm N}$ determined from the specific heat measurements, the peak value of $C/T$ at $T_{\mathrm N}$, and $\kappa_{xx}^{\mathrm{peak}}$. Although the precise values of these quantities may depend slightly on the measurement techniques used, the high values obtained in our samples confirm the superior quality of crystals grown by the present method. A more comprehensive study of the thermal transport properties under magnetic fields on the high-quality single crystals synthesized in this study will address issues regarding the reproducibility of the half-integer quantum thermal Hall effect in $\kappa_{xy}$ and the origin of the oscillations in $\kappa_{xx}$ under magnetic fields.

\begin{table}[t]
\centering
\caption{$T_{\mathrm{N}}$, specific heat $C/T$ at $T_{\mathrm{N}}$, and $\kappa_{xx}^{\mathrm{peak}}$ for samples synthesized by different crystal growth methods. Note that in Refs.~\cite{Kasahara2022,Tanaka2022,imamura2023irradiation}, specific heat was measured using the same long-relaxation method as in this study, while in Refs.~\cite{Zhang2024,Widmann2019}, specific heat was measured using the Quantum Design Physical Property Measurement System (PPMS).}
\begin{tabular}{@{\hspace{1pt}}c@{\hspace{10pt}}c@{\hspace{10pt}}c@{\hspace{10pt}}c} \hline
   Growth method & \textit{T}$_{\mathrm{N}}$ & \multicolumn{1}{c}{\textit{C/T} at \textit{T}$_{\mathrm{N}}$} & \multicolumn{1}{c}{$\kappa_{xx}^{\mathrm{peak}}$} \\
    & (K) & (J$\mathrm{mol^{-1}}$$\mathrm{K^{-2}}$) & ($\mathrm{W/Km}$) \\ \hline \hline
   Two-step sublimation & 7.6-7.7 & 1.5 & 13 \\
   SSVG~\cite{Zhang2024} & 7.6 & 2.0 & 8 \\
   Bridgman~\cite{Kasahara2022} & 7.4 & 0.7 & 4 \\
   Bridgman~\cite{Tanaka2022} & 7.0 & 0.6 & - \\
   One-step sublimation~\cite{Widmann2019} & 6.5 & 0.5 & - \\
   Bridgman (irradiated)~\cite{imamura2023irradiation} & 5.7 & 0.2 & - \\ \hline
\end{tabular}
\label{Table1}
\end{table}

\section{Conclusion}
In summary, we synthesized high-quality single crystals of $\alpha$-RuCl$_{3}$ by the two-step sublimation method and characterized their physical properties. The magnetic susceptibility and thermal expansion coefficient show a distinct hysteresis loop at $\sim 150$\,K, indicating the clear first-order transition at $T_{\mathrm s}$. The $^{35}$Cl NQR spectra demonstrate a crystal structure change from monoclinic ($C2/m$) to rhombohedral ($R\bar{3}$) at the structural phase transition. The Raman spectra show only the phonon peaks expected in an isolated RuCl$_{3}$ layer, supporting the good crystallinity of our samples. The jumps in the thermal expansion coefficient and specific heat at $T_{\mathrm N}$ are much higher and sharper than those of previous samples grown by the CVT and Bridgman methods and do not show any additional AFM transitions at 10-14\,K due to stacking faults. The longitudinal thermal conductivity in the AFM phase is significantly larger than previously reported. All the results indicate that our single crystals are of high quality with good crystallinity and few stacking faults, which provide a platform for resolving the discrepancies discussed in the thermal transport measurements on $\alpha$-RuCl$_{3}$.

\section{Acknowledgements}
We thank T.~Kurumaji, R.~Ishii, S.~Dekura, S.~Kitou, H.~Mori, Y.~Uwatoko, and J.~M$\ddot{\mathrm{u}}$ller for fruitful discussions and technical supports. This work was supported by CREST (JPMJCR19T5) from Japan Science and Technology (JST), Grants-in-Aid for Scientific Research (KAKENHI) (Grants Nos.\,JP24K17007, JP24H01646, JP23H00089, JP22H00105, JP22K14005, JP22K18681, JP22K18683, JP21H01793, JP21H04988, JP19H00649, JP18H05227, and JP18KK0375), Grant-in-Aid for Scientific Research on Innovative Areas ``Quantum Liquid Crystals” (No.\,JP19H05824), Transformative Research Areas (A) ``Condensed Conjugation” (Nos.\,JP23H04025, JP20H05869) and ``Extreme Universe” (No.\,JP22H05256) from Japan Society for the Promotion of Science, and the Deutsche Forschungsgemeinschaft (DFG, German Research Foundation) through TRR 288-422213477 (Project A06).

\section{Appendix}
\begin{figure}[t]
    \includegraphics[width=1\linewidth]{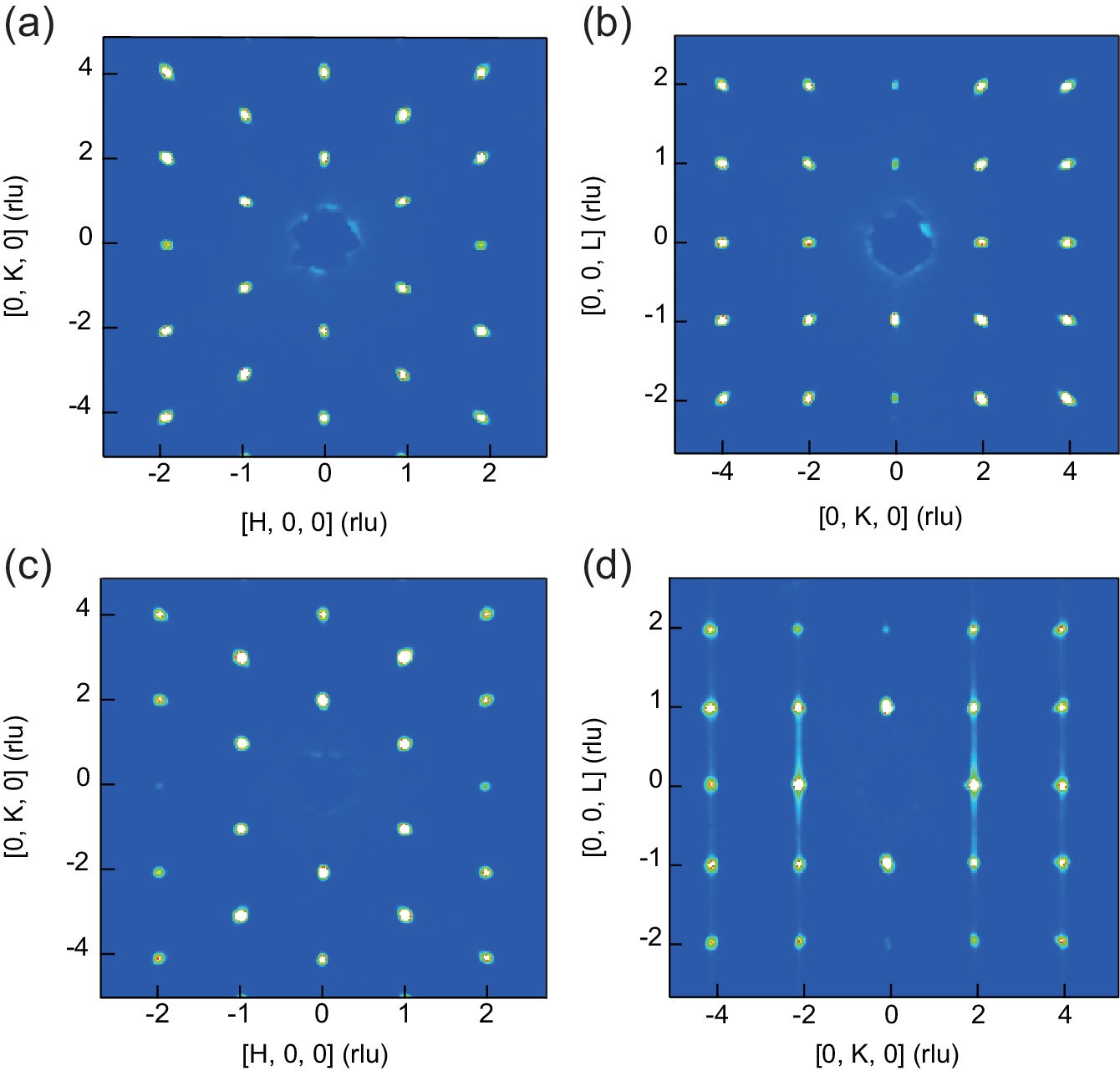}
    \caption{(a), (b) XRD patterns in the $(H, K, 0)$ and $(0, K, L)$ planes for the as-grown sample (\#X1), respectively. (c), (d) XRD patterns in the $(H, K, 0)$ and $(0, K, L)$ planes for the cut sample (\#X2), respectively.
    }
    \label{Fig6}
\end{figure}

\begin{figure}[t]
    \includegraphics[width=0.9\linewidth]{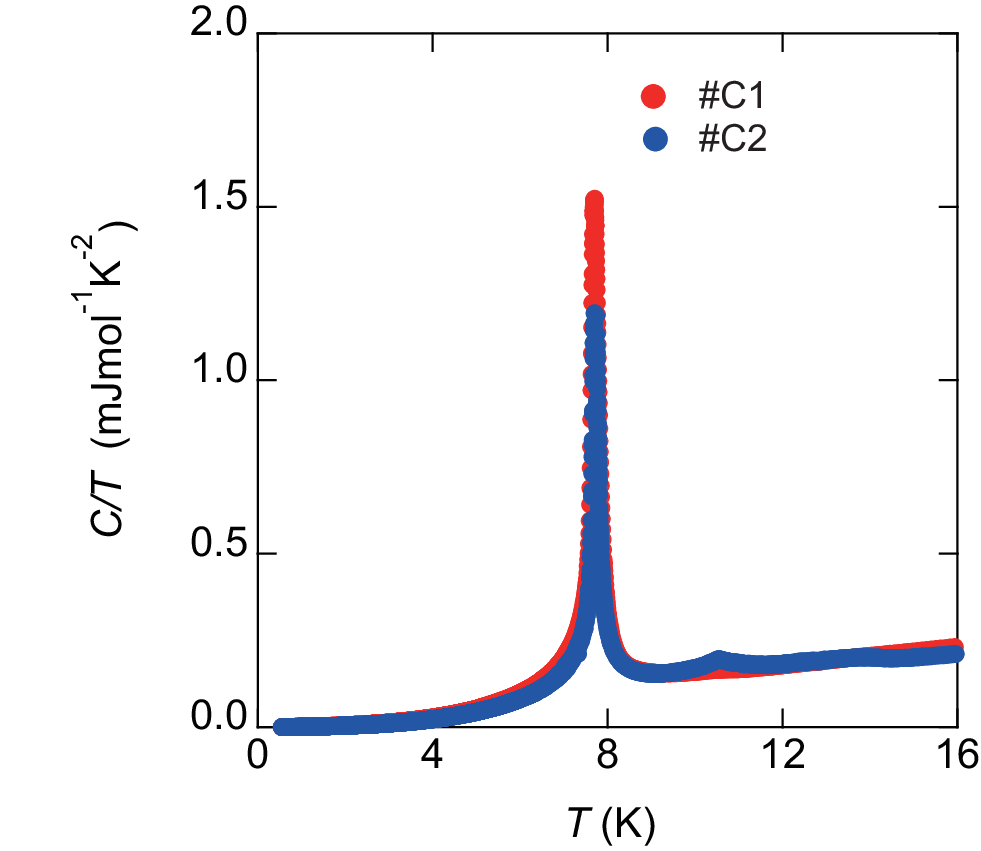}
    \caption{Temperature dependence of specific heat divided by temperature $C/T$ for the as-grown sample (red circles, \#C1) and the cut sample (blue circles, \#C2).
    }
    \label{Fig7}
\end{figure}

We performed XRD measurements at room temperature to check the crystallinity of $\alpha$-RuCl$_{3}$ single crystals obtained in this study. Figure\,6(a) and 6(b) show the XRD patterns in the $(H, K, 0)$ and $(0, K, L)$ planes for an as-grown sample, respectively. 
The diffraction patterns in the $(H, K, 0)$ and $(0, K, L)$ planes both show clean spots, indicating good crystallinity in both the $ab$ plane and the stacking ($c$-axis) direction. Here, we stress that the $(0, K, L)$ plane diffraction, which includes information on the stacking direction, shows no diffuse lines but clean spots, indicating minimal stacking faults in our as-grown single crystal. Furthermore, we performed XRD measurements for a sample after cutting with a razor blade. As a result, we have found that while the XRD spots in the $(H, K, 0)$ plane do not change with the cut, diffusive lines along the stacking direction appear in the $(0, K, L)$ plane [see Figs.\,6(c) and 6(d)]. These results indicate that when the sample is cut, few defects are introduced in the $ab$ plane, while many stacking faults are introduced in the stacking direction. Correspondingly, slight anomalies appear in the specific heat $C/T$ at around 10\,K and 14\,K (see Fig.\,7). Therefore, the clear spots in the XRD patterns and the absence of anomalies in the specific heat for the as-grown samples indicate minimal stacking faults in our single crystals.

\bibliography{sublimation_ref}

\begin{thebibliography}{46}%
\makeatletter
\providecommand \@ifxundefined [1]{%
 \@ifx{#1\undefined}
}%
\providecommand \@ifnum [1]{%
 \ifnum #1\expandafter \@firstoftwo
 \else \expandafter \@secondoftwo
 \fi
}%
\providecommand \@ifx [1]{%
 \ifx #1\expandafter \@firstoftwo
 \else \expandafter \@secondoftwo
 \fi
}%
\providecommand \natexlab [1]{#1}%
\providecommand \enquote  [1]{``#1''}%
\providecommand \bibnamefont  [1]{#1}%
\providecommand \bibfnamefont [1]{#1}%
\providecommand \citenamefont [1]{#1}%
\providecommand \href@noop [0]{\@secondoftwo}%
\providecommand \href [0]{\begingroup \@sanitize@url \@href}%
\providecommand \@href[1]{\@@startlink{#1}\@@href}%
\providecommand \@@href[1]{\endgroup#1\@@endlink}%
\providecommand \@sanitize@url [0]{\catcode `\\12\catcode `\$12\catcode
  `\&12\catcode `\#12\catcode `\^12\catcode `\_12\catcode `\%12\relax}%
\providecommand \@@startlink[1]{}%
\providecommand \@@endlink[0]{}%
\providecommand \url  [0]{\begingroup\@sanitize@url \@url }%
\providecommand \@url [1]{\endgroup\@href {#1}{\urlprefix }}%
\providecommand \urlprefix  [0]{URL }%
\providecommand \Eprint [0]{\href }%
\providecommand \doibase [0]{https://doi.org/}%
\providecommand \selectlanguage [0]{\@gobble}%
\providecommand \bibinfo  [0]{\@secondoftwo}%
\providecommand \bibfield  [0]{\@secondoftwo}%
\providecommand \translation [1]{[#1]}%
\providecommand \BibitemOpen [0]{}%
\providecommand \bibitemStop [0]{}%
\providecommand \bibitemNoStop [0]{.\EOS\space}%
\providecommand \EOS [0]{\spacefactor3000\relax}%
\providecommand \BibitemShut  [1]{\csname bibitem#1\endcsname}%
\let\auto@bib@innerbib\@empty
\bibitem [{\citenamefont {Balents}(2010)}]{Balents2010}%
  \BibitemOpen
  \bibfield  {author} {\bibinfo {author} {\bibfnamefont {L.}~\bibnamefont
  {Balents}},\ }\bibfield  {title} {\bibinfo {title} {Spin liquids in
  frustrated magnets},\ }\href {https://doi.org/10.1038/nature08917} {\bibfield
   {journal} {\bibinfo  {journal} {{Nature}}\ }\textbf {\bibinfo {volume}
  {464}},\ \bibinfo {pages} {199} (\bibinfo {year} {2010})}\BibitemShut
  {NoStop}%
\bibitem [{\citenamefont {Anderson}(1973)}]{Anderson1973}%
  \BibitemOpen
  \bibfield  {author} {\bibinfo {author} {\bibfnamefont {P.~W.}\ \bibnamefont
  {Anderson}},\ }\bibfield  {title} {\bibinfo {title} {Resonating valence
  bonds: A new kind of insulator?},\ }\href
  {https://doi.org/http://dx.doi.org/10.1016/0025-5408(73)90167-0} {\bibfield
  {journal} {\bibinfo  {journal} {Mater. Res. Bull.}\ }\textbf {\bibinfo
  {volume} {8}},\ \bibinfo {pages} {153} (\bibinfo {year} {1973})}\BibitemShut
  {NoStop}%
\bibitem [{\citenamefont {{Kitaev}}(2006)}]{Kitaev2006}%
  \BibitemOpen
  \bibfield  {author} {\bibinfo {author} {\bibfnamefont {A.}~\bibnamefont
  {{Kitaev}}},\ }\bibfield  {title} {\bibinfo {title} {Anyons in an exactly
  solved model and beyond},\ }\href {https://doi.org/10.1016/j.aop.2005.10.005}
  {\bibfield  {journal} {\bibinfo  {journal} {Ann. Phys.}\ }\textbf {\bibinfo
  {volume} {321}},\ \bibinfo {pages} {2} (\bibinfo {year} {2006})}\BibitemShut
  {NoStop}%
\bibitem [{\citenamefont {Motome}\ and\ \citenamefont
  {Nasu}(2020)}]{Motome2020}%
  \BibitemOpen
  \bibfield  {author} {\bibinfo {author} {\bibfnamefont {Y.}~\bibnamefont
  {Motome}}\ and\ \bibinfo {author} {\bibfnamefont {J.}~\bibnamefont {Nasu}},\
  }\bibfield  {title} {\bibinfo {title} {Hunting {Majorana} fermions in
  {Kitaev} magnets},\ }\href {https://doi.org/10.7566/JPSJ.89.012002}
  {\bibfield  {journal} {\bibinfo  {journal} {J. Phys. Soc. Jpn.}\ }\textbf
  {\bibinfo {volume} {89}},\ \bibinfo {pages} {012002} (\bibinfo {year}
  {2020})}\BibitemShut {NoStop}%
\bibitem [{\citenamefont {Jackeli}\ and\ \citenamefont
  {Khaliullin}(2009)}]{Jackeli2009}%
  \BibitemOpen
  \bibfield  {author} {\bibinfo {author} {\bibfnamefont {G.}~\bibnamefont
  {Jackeli}}\ and\ \bibinfo {author} {\bibfnamefont {G.}~\bibnamefont
  {Khaliullin}},\ }\bibfield  {title} {\bibinfo {title} {Mott insulators in the
  strong spin-orbit coupling limit: From $\rm{Heisenberg}$ to a quantum compass
  and $\rm{Kitaev}$ models},\ }\href
  {https://doi.org/10.1103/PhysRevLett.102.017205} {\bibfield  {journal}
  {\bibinfo  {journal} {Phys. Rev. Lett.}\ }\textbf {\bibinfo {volume} {102}},\
  \bibinfo {pages} {017205} (\bibinfo {year} {2009})}\BibitemShut {NoStop}%
\bibitem [{\citenamefont {Takagi}\ \emph {et~al.}(2019)\citenamefont {Takagi},
  \citenamefont {Takayama}, \citenamefont {Jackeli}, \citenamefont
  {Khaliullin},\ and\ \citenamefont {Nagler}}]{Takagi2019}%
  \BibitemOpen
  \bibfield  {author} {\bibinfo {author} {\bibfnamefont {H.}~\bibnamefont
  {Takagi}}, \bibinfo {author} {\bibfnamefont {T.}~\bibnamefont {Takayama}},
  \bibinfo {author} {\bibfnamefont {G.}~\bibnamefont {Jackeli}}, \bibinfo
  {author} {\bibfnamefont {G.}~\bibnamefont {Khaliullin}},\ and\ \bibinfo
  {author} {\bibfnamefont {S.~E.}\ \bibnamefont {Nagler}},\ }\bibfield  {title}
  {\bibinfo {title} {{Concept and realization of {Kitaev} quantum spin
  liquids}},\ }\href {https://doi.org/10.1038/s42254-019-0038-2} {\bibfield
  {journal} {\bibinfo  {journal} {Nat. Rev. Phys.}\ }\textbf {\bibinfo {volume}
  {1}},\ \bibinfo {pages} {264} (\bibinfo {year} {2019})}\BibitemShut {NoStop}%
\bibitem [{\citenamefont {Plumb}\ \emph {et~al.}(2014)\citenamefont {Plumb},
  \citenamefont {Clancy}, \citenamefont {Sandilands}, \citenamefont {Shankar},
  \citenamefont {Hu}, \citenamefont {Burch}, \citenamefont {Kee},\ and\
  \citenamefont {Kim}}]{Plumb2014}%
  \BibitemOpen
  \bibfield  {author} {\bibinfo {author} {\bibfnamefont {K.~W.}\ \bibnamefont
  {Plumb}}, \bibinfo {author} {\bibfnamefont {J.~P.}\ \bibnamefont {Clancy}},
  \bibinfo {author} {\bibfnamefont {L.~J.}\ \bibnamefont {Sandilands}},
  \bibinfo {author} {\bibfnamefont {V.~V.}\ \bibnamefont {Shankar}}, \bibinfo
  {author} {\bibfnamefont {Y.~F.}\ \bibnamefont {Hu}}, \bibinfo {author}
  {\bibfnamefont {K.~S.}\ \bibnamefont {Burch}}, \bibinfo {author}
  {\bibfnamefont {H.-Y.}\ \bibnamefont {Kee}},\ and\ \bibinfo {author}
  {\bibfnamefont {Y.-J.}\ \bibnamefont {Kim}},\ }\bibfield  {title} {\bibinfo
  {title} {{$\alpha$-{RuCl}$_{3}$: A spin-orbit assisted {Mott} insulator on a
  honeycomb lattice}},\ }\href {https://doi.org/10.1103/PhysRevB.90.041112}
  {\bibfield  {journal} {\bibinfo  {journal} {Phys. Rev. B}\ }\textbf {\bibinfo
  {volume} {90}},\ \bibinfo {pages} {041112} (\bibinfo {year}
  {2014})}\BibitemShut {NoStop}%
\bibitem [{\citenamefont {Kubota}\ \emph {et~al.}(2015)\citenamefont {Kubota},
  \citenamefont {Tanaka}, \citenamefont {Ono}, \citenamefont {Narumi},\ and\
  \citenamefont {Kindo}}]{Kubota2015}%
  \BibitemOpen
  \bibfield  {author} {\bibinfo {author} {\bibfnamefont {Y.}~\bibnamefont
  {Kubota}}, \bibinfo {author} {\bibfnamefont {H.}~\bibnamefont {Tanaka}},
  \bibinfo {author} {\bibfnamefont {T.}~\bibnamefont {Ono}}, \bibinfo {author}
  {\bibfnamefont {Y.}~\bibnamefont {Narumi}},\ and\ \bibinfo {author}
  {\bibfnamefont {K.}~\bibnamefont {Kindo}},\ }\bibfield  {title} {\bibinfo
  {title} {{Successive magnetic phase transitions in $\alpha$-RuCl$_3$:
  $XY$-like frustrated magnet on the honeycomb lattice}},\ }\href
  {https://doi.org/10.1103/PhysRevB.91.094422} {\bibfield  {journal} {\bibinfo
  {journal} {Phys. Rev. B}\ }\textbf {\bibinfo {volume} {91}},\ \bibinfo
  {pages} {094422} (\bibinfo {year} {2015})}\BibitemShut {NoStop}%
\bibitem [{\citenamefont {Majumder}\ \emph {et~al.}(2015)\citenamefont
  {Majumder}, \citenamefont {Schmidt}, \citenamefont {Rosner}, \citenamefont
  {Tsirlin}, \citenamefont {Yasuoka},\ and\ \citenamefont
  {Baenitz}}]{Majumder2015}%
  \BibitemOpen
  \bibfield  {author} {\bibinfo {author} {\bibfnamefont {M.}~\bibnamefont
  {Majumder}}, \bibinfo {author} {\bibfnamefont {M.}~\bibnamefont {Schmidt}},
  \bibinfo {author} {\bibfnamefont {H.}~\bibnamefont {Rosner}}, \bibinfo
  {author} {\bibfnamefont {A.~A.}\ \bibnamefont {Tsirlin}}, \bibinfo {author}
  {\bibfnamefont {H.}~\bibnamefont {Yasuoka}},\ and\ \bibinfo {author}
  {\bibfnamefont {M.}~\bibnamefont {Baenitz}},\ }\bibfield  {title} {\bibinfo
  {title} {{Anisotropic Ru$^{3+}$ 4{\it d}$^5$ magnetism in the
  $\alpha$-RuCl$_3$ honeycomb system: Susceptibility, specific heat, and
  zero-field NMR}},\ }\href {https://doi.org/10.1103/PhysRevB.91.180401}
  {\bibfield  {journal} {\bibinfo  {journal} {Phys. Rev. B}\ }\textbf {\bibinfo
  {volume} {91}},\ \bibinfo {pages} {180401} (\bibinfo {year}
  {2015})}\BibitemShut {NoStop}%
\bibitem [{\citenamefont {Johnson}\ \emph {et~al.}(2015)\citenamefont
  {Johnson}, \citenamefont {Williams}, \citenamefont {Haghighirad},
  \citenamefont {Singleton}, \citenamefont {Zapf}, \citenamefont {Manuel},
  \citenamefont {Mazin}, \citenamefont {Li}, \citenamefont {Jeschke},
  \citenamefont {Valent\'{\i}},\ and\ \citenamefont {Coldea}}]{Johnson2015}%
  \BibitemOpen
  \bibfield  {author} {\bibinfo {author} {\bibfnamefont {R.~D.}\ \bibnamefont
  {Johnson}}, \bibinfo {author} {\bibfnamefont {S.~C.}\ \bibnamefont
  {Williams}}, \bibinfo {author} {\bibfnamefont {A.~A.}\ \bibnamefont
  {Haghighirad}}, \bibinfo {author} {\bibfnamefont {J.}~\bibnamefont
  {Singleton}}, \bibinfo {author} {\bibfnamefont {V.}~\bibnamefont {Zapf}},
  \bibinfo {author} {\bibfnamefont {P.}~\bibnamefont {Manuel}}, \bibinfo
  {author} {\bibfnamefont {I.~I.}\ \bibnamefont {Mazin}}, \bibinfo {author}
  {\bibfnamefont {Y.}~\bibnamefont {Li}}, \bibinfo {author} {\bibfnamefont
  {H.~O.}\ \bibnamefont {Jeschke}}, \bibinfo {author} {\bibfnamefont
  {R.}~\bibnamefont {Valent\'{\i}}},\ and\ \bibinfo {author} {\bibfnamefont
  {R.}~\bibnamefont {Coldea}},\ }\bibfield  {title} {\bibinfo {title}
  {Monoclinic crystal structure of $\alpha$-{RuCl}$_{3}$ and the zigzag
  antiferromagnetic ground state},\ }\href
  {https://doi.org/10.1103/PhysRevB.92.235119} {\bibfield  {journal} {\bibinfo
  {journal} {Phys. Rev. B}\ }\textbf {\bibinfo {volume} {92}},\ \bibinfo
  {pages} {235119} (\bibinfo {year} {2015})}\BibitemShut {NoStop}%
\bibitem [{\citenamefont {Winter}\ \emph {et~al.}(2016)\citenamefont {Winter},
  \citenamefont {Li}, \citenamefont {Jeschke},\ and\ \citenamefont
  {Valent\'{\i}}}]{Winter2016}%
  \BibitemOpen
  \bibfield  {author} {\bibinfo {author} {\bibfnamefont {S.~M.}\ \bibnamefont
  {Winter}}, \bibinfo {author} {\bibfnamefont {Y.}~\bibnamefont {Li}}, \bibinfo
  {author} {\bibfnamefont {H.~O.}\ \bibnamefont {Jeschke}},\ and\ \bibinfo
  {author} {\bibfnamefont {R.}~\bibnamefont {Valent\'{\i}}},\ }\bibfield
  {title} {\bibinfo {title} {{Challenges in design of Kitaev materials:
  Magnetic interactions from competing energy scales}},\ }\href
  {https://doi.org/10.1103/PhysRevB.93.214431} {\bibfield  {journal} {\bibinfo
  {journal} {Phys. Rev. B}\ }\textbf {\bibinfo {volume} {93}},\ \bibinfo
  {pages} {214431} (\bibinfo {year} {2016})}\BibitemShut {NoStop}%
\bibitem [{\citenamefont {Banerjee}\ \emph {et~al.}(2016)\citenamefont
  {Banerjee}, \citenamefont {Bridges}, \citenamefont {Yan}, \citenamefont
  {Aczel}, \citenamefont {Li}, \citenamefont {Stone}, \citenamefont {Granroth},
  \citenamefont {Lumsden}, \citenamefont {Yiu}, \citenamefont {Knolle} \emph
  {et~al.}}]{Banerjee2016proximate}%
  \BibitemOpen
  \bibfield  {author} {\bibinfo {author} {\bibfnamefont {A.}~\bibnamefont
  {Banerjee}}, \bibinfo {author} {\bibfnamefont {C.}~\bibnamefont {Bridges}},
  \bibinfo {author} {\bibfnamefont {J.-Q.}\ \bibnamefont {Yan}}, \bibinfo
  {author} {\bibfnamefont {A.}~\bibnamefont {Aczel}}, \bibinfo {author}
  {\bibfnamefont {L.}~\bibnamefont {Li}}, \bibinfo {author} {\bibfnamefont
  {M.}~\bibnamefont {Stone}}, \bibinfo {author} {\bibfnamefont
  {G.}~\bibnamefont {Granroth}}, \bibinfo {author} {\bibfnamefont
  {M.}~\bibnamefont {Lumsden}}, \bibinfo {author} {\bibfnamefont
  {Y.}~\bibnamefont {Yiu}}, \bibinfo {author} {\bibfnamefont {J.}~\bibnamefont
  {Knolle}}, \emph {et~al.},\ }\bibfield  {title} {\bibinfo {title} {{Proximate
  Kitaev quantum spin liquid behaviour in a honeycomb magnet}},\ }\href
  {https://doi.org/10.1038/nmat4604} {\bibfield  {journal} {\bibinfo  {journal}
  {Nat. Mater.}\ }\textbf {\bibinfo {volume} {15}},\ \bibinfo {pages} {733}
  (\bibinfo {year} {2016})}\BibitemShut {NoStop}%
\bibitem [{\citenamefont {Banerjee}\ \emph {et~al.}(2017)\citenamefont
  {Banerjee}, \citenamefont {Yan}, \citenamefont {Knolle}, \citenamefont
  {Bridges}, \citenamefont {Stone}, \citenamefont {Lumsden}, \citenamefont
  {Mandrus}, \citenamefont {Tennant}, \citenamefont {Moessner},\ and\
  \citenamefont {Nagler}}]{Banerjee2017}%
  \BibitemOpen
  \bibfield  {author} {\bibinfo {author} {\bibfnamefont {A.}~\bibnamefont
  {Banerjee}}, \bibinfo {author} {\bibfnamefont {J.}~\bibnamefont {Yan}},
  \bibinfo {author} {\bibfnamefont {J.}~\bibnamefont {Knolle}}, \bibinfo
  {author} {\bibfnamefont {C.~A.}\ \bibnamefont {Bridges}}, \bibinfo {author}
  {\bibfnamefont {M.~B.}\ \bibnamefont {Stone}}, \bibinfo {author}
  {\bibfnamefont {M.~D.}\ \bibnamefont {Lumsden}}, \bibinfo {author}
  {\bibfnamefont {D.~G.}\ \bibnamefont {Mandrus}}, \bibinfo {author}
  {\bibfnamefont {D.~A.}\ \bibnamefont {Tennant}}, \bibinfo {author}
  {\bibfnamefont {R.}~\bibnamefont {Moessner}},\ and\ \bibinfo {author}
  {\bibfnamefont {S.~E.}\ \bibnamefont {Nagler}},\ }\bibfield  {title}
  {\bibinfo {title} {{Neutron scattering in the proximate quantum spin liquid
  $\alpha$-RuCl$_3$}},\ }\href
  {https://www.science.org/doi/abs/10.1126/science.aah6015} {\bibfield
  {journal} {\bibinfo  {journal} {Science}\ }\textbf {\bibinfo {volume}
  {356}},\ \bibinfo {pages} {1055} (\bibinfo {year} {2017})}\BibitemShut
  {NoStop}%
\bibitem [{\citenamefont {Do}\ \emph {et~al.}(2017)\citenamefont {Do},
  \citenamefont {Park}, \citenamefont {Yoshitake}, \citenamefont {Nasu},
  \citenamefont {Motome}, \citenamefont {Kwon}, \citenamefont {Adroja},
  \citenamefont {Voneshen}, \citenamefont {Kim}, \citenamefont {Jang},
  \citenamefont {Park}, \citenamefont {Choi},\ and\ \citenamefont
  {Ji}}]{Do2017}%
  \BibitemOpen
  \bibfield  {author} {\bibinfo {author} {\bibfnamefont {S.-H.}\ \bibnamefont
  {Do}}, \bibinfo {author} {\bibfnamefont {S.-Y.}\ \bibnamefont {Park}},
  \bibinfo {author} {\bibfnamefont {J.}~\bibnamefont {Yoshitake}}, \bibinfo
  {author} {\bibfnamefont {J.}~\bibnamefont {Nasu}}, \bibinfo {author}
  {\bibfnamefont {Y.}~\bibnamefont {Motome}}, \bibinfo {author} {\bibfnamefont
  {Y.~S.}\ \bibnamefont {Kwon}}, \bibinfo {author} {\bibfnamefont {D.~T.}\
  \bibnamefont {Adroja}}, \bibinfo {author} {\bibfnamefont {D.~J.}\
  \bibnamefont {Voneshen}}, \bibinfo {author} {\bibfnamefont {K.}~\bibnamefont
  {Kim}}, \bibinfo {author} {\bibfnamefont {T.-H.}\ \bibnamefont {Jang}},
  \bibinfo {author} {\bibfnamefont {J.-H.}\ \bibnamefont {Park}}, \bibinfo
  {author} {\bibfnamefont {K.-Y.}\ \bibnamefont {Choi}},\ and\ \bibinfo
  {author} {\bibfnamefont {S.}~\bibnamefont {Ji}},\ }\bibfield  {title}
  {\bibinfo {title} {{Majorana} fermions in the {Kitaev} quantum spin system
  $\alpha$-{RuCl}$_3$},\ }\href {https://doi.org/10.1038/nphys4264} {\bibfield
  {journal} {\bibinfo  {journal} {Nat. Phys.}\ }\textbf {\bibinfo {volume}
  {13}},\ \bibinfo {pages} {1079} (\bibinfo {year} {2017})}\BibitemShut
  {NoStop}%
\bibitem [{\citenamefont {Sandilands}\ \emph {et~al.}(2015)\citenamefont
  {Sandilands}, \citenamefont {Tian}, \citenamefont {Plumb}, \citenamefont
  {Kim},\ and\ \citenamefont {Burch}}]{Sandilands2015}%
  \BibitemOpen
  \bibfield  {author} {\bibinfo {author} {\bibfnamefont {L.~J.}\ \bibnamefont
  {Sandilands}}, \bibinfo {author} {\bibfnamefont {Y.}~\bibnamefont {Tian}},
  \bibinfo {author} {\bibfnamefont {K.~W.}\ \bibnamefont {Plumb}}, \bibinfo
  {author} {\bibfnamefont {Y.-J.}\ \bibnamefont {Kim}},\ and\ \bibinfo {author}
  {\bibfnamefont {K.~S.}\ \bibnamefont {Burch}},\ }\bibfield  {title} {\bibinfo
  {title} {{Scattering continuum and possible fractionalized excitations in
  $\alpha$-RuCl$_3$}},\ }\href {https://doi.org/10.1103/PhysRevLett.114.147201}
  {\bibfield  {journal} {\bibinfo  {journal} {Phys. Rev. Lett.}\ }\textbf
  {\bibinfo {volume} {114}},\ \bibinfo {pages} {147201} (\bibinfo {year}
  {2015})}\BibitemShut {NoStop}%
\bibitem [{\citenamefont {Nasu}\ \emph {et~al.}(2016)\citenamefont {Nasu},
  \citenamefont {Knolle}, \citenamefont {Kovrizhin}, \citenamefont {Motome},\
  and\ \citenamefont {Moessner}}]{Nasu2016}%
  \BibitemOpen
  \bibfield  {author} {\bibinfo {author} {\bibfnamefont {J.}~\bibnamefont
  {Nasu}}, \bibinfo {author} {\bibfnamefont {J.}~\bibnamefont {Knolle}},
  \bibinfo {author} {\bibfnamefont {D.~L.}\ \bibnamefont {Kovrizhin}}, \bibinfo
  {author} {\bibfnamefont {Y.}~\bibnamefont {Motome}},\ and\ \bibinfo {author}
  {\bibfnamefont {R.}~\bibnamefont {Moessner}},\ }\bibfield  {title} {\bibinfo
  {title} {Fermionic response from fractionalization in an insulating
  two-dimensional magnet},\ }\href {https://doi.org/10.1038/nphys3809}
  {\bibfield  {journal} {\bibinfo  {journal} {Nat. Phys.}\ }\textbf {\bibinfo
  {volume} {12}},\ \bibinfo {pages} {912} (\bibinfo {year} {2016})}\BibitemShut
  {NoStop}%
\bibitem [{\citenamefont {Widmann}\ \emph {et~al.}(2019)\citenamefont
  {Widmann}, \citenamefont {Tsurkan}, \citenamefont {Prishchenko},
  \citenamefont {Mazurenko}, \citenamefont {Tsirlin},\ and\ \citenamefont
  {Loidl}}]{Widmann2019}%
  \BibitemOpen
  \bibfield  {author} {\bibinfo {author} {\bibfnamefont {S.}~\bibnamefont
  {Widmann}}, \bibinfo {author} {\bibfnamefont {V.}~\bibnamefont {Tsurkan}},
  \bibinfo {author} {\bibfnamefont {D.~A.}\ \bibnamefont {Prishchenko}},
  \bibinfo {author} {\bibfnamefont {V.~G.}\ \bibnamefont {Mazurenko}}, \bibinfo
  {author} {\bibfnamefont {A.~A.}\ \bibnamefont {Tsirlin}},\ and\ \bibinfo
  {author} {\bibfnamefont {A.}~\bibnamefont {Loidl}},\ }\bibfield  {title}
  {\bibinfo {title} {{Thermodynamic evidence of fractionalized excitations in
  $\alpha$-RuCl$_3$}},\ }\href {https://doi.org/10.1103/PhysRevB.99.094415}
  {\bibfield  {journal} {\bibinfo  {journal} {Phys. Rev. B}\ }\textbf {\bibinfo
  {volume} {99}},\ \bibinfo {pages} {094415} (\bibinfo {year}
  {2019})}\BibitemShut {NoStop}%
\bibitem [{\citenamefont {Yadav}\ \emph {et~al.}(2016)\citenamefont {Yadav},
  \citenamefont {Bogdanov}, \citenamefont {Katukuri}, \citenamefont
  {Nishimoto}, \citenamefont {van~den Brink},\ and\ \citenamefont
  {Hozoi}}]{Yadav2016}%
  \BibitemOpen
  \bibfield  {author} {\bibinfo {author} {\bibfnamefont {R.}~\bibnamefont
  {Yadav}}, \bibinfo {author} {\bibfnamefont {N.~A.}\ \bibnamefont {Bogdanov}},
  \bibinfo {author} {\bibfnamefont {V.~M.}\ \bibnamefont {Katukuri}}, \bibinfo
  {author} {\bibfnamefont {S.}~\bibnamefont {Nishimoto}}, \bibinfo {author}
  {\bibfnamefont {J.}~\bibnamefont {van~den Brink}},\ and\ \bibinfo {author}
  {\bibfnamefont {L.}~\bibnamefont {Hozoi}},\ }\bibfield  {title} {\bibinfo
  {title} {{Kitaev} exchange and field-induced quantum spin-liquid states in
  honeycomb $\alpha$-{RuCl}$_{3}$},\ }\href {https://doi.org/10.1038/srep37925}
  {\bibfield  {journal} {\bibinfo  {journal} {Sci. Rep.}\ }\textbf {\bibinfo
  {volume} {6}},\ \bibinfo {pages} {37925} (\bibinfo {year}
  {2016})}\BibitemShut {NoStop}%
\bibitem [{\citenamefont {Wolter}\ \emph {et~al.}(2017)\citenamefont {Wolter},
  \citenamefont {Corredor}, \citenamefont {Janssen}, \citenamefont {Nenkov},
  \citenamefont {Sch\"onecker}, \citenamefont {Do}, \citenamefont {Choi},
  \citenamefont {Albrecht}, \citenamefont {Hunger}, \citenamefont {Doert},
  \citenamefont {Vojta},\ and\ \citenamefont {B\"uchner}}]{Wolter2017}%
  \BibitemOpen
  \bibfield  {author} {\bibinfo {author} {\bibfnamefont {A.~U.~B.}\
  \bibnamefont {Wolter}}, \bibinfo {author} {\bibfnamefont {L.~T.}\
  \bibnamefont {Corredor}}, \bibinfo {author} {\bibfnamefont {L.}~\bibnamefont
  {Janssen}}, \bibinfo {author} {\bibfnamefont {K.}~\bibnamefont {Nenkov}},
  \bibinfo {author} {\bibfnamefont {S.}~\bibnamefont {Sch\"onecker}}, \bibinfo
  {author} {\bibfnamefont {S.-H.}\ \bibnamefont {Do}}, \bibinfo {author}
  {\bibfnamefont {K.-Y.}\ \bibnamefont {Choi}}, \bibinfo {author}
  {\bibfnamefont {R.}~\bibnamefont {Albrecht}}, \bibinfo {author}
  {\bibfnamefont {J.}~\bibnamefont {Hunger}}, \bibinfo {author} {\bibfnamefont
  {T.}~\bibnamefont {Doert}}, \bibinfo {author} {\bibfnamefont
  {M.}~\bibnamefont {Vojta}},\ and\ \bibinfo {author} {\bibfnamefont
  {B.}~\bibnamefont {B\"uchner}},\ }\bibfield  {title} {\bibinfo {title}
  {{Field-induced quantum criticality in the $\rm{Kitaev}$ system
  $\alpha$-RuCl$_3$}},\ }\href {https://doi.org/10.1103/PhysRevB.96.041405}
  {\bibfield  {journal} {\bibinfo  {journal} {Phys. Rev. B}\ }\textbf {\bibinfo
  {volume} {96}},\ \bibinfo {pages} {041405} (\bibinfo {year}
  {2017})}\BibitemShut {NoStop}%
\bibitem [{\citenamefont {Banerjee}\ \emph {et~al.}(2018)\citenamefont
  {Banerjee}, \citenamefont {Lampen-Kelley}, \citenamefont {Knolle},
  \citenamefont {Balz}, \citenamefont {Aczel}, \citenamefont {Winn},
  \citenamefont {Liu}, \citenamefont {Pajerowski}, \citenamefont {Yan},
  \citenamefont {Bridges}, \citenamefont {Savici}, \citenamefont {Chakoumakos},
  \citenamefont {Lumsden}, \citenamefont {Tennant}, \citenamefont {Moessner},
  \citenamefont {Mandrus},\ and\ \citenamefont {Nagler}}]{Banerjee2018}%
  \BibitemOpen
  \bibfield  {author} {\bibinfo {author} {\bibfnamefont {A.}~\bibnamefont
  {Banerjee}}, \bibinfo {author} {\bibfnamefont {P.}~\bibnamefont
  {Lampen-Kelley}}, \bibinfo {author} {\bibfnamefont {J.}~\bibnamefont
  {Knolle}}, \bibinfo {author} {\bibfnamefont {C.}~\bibnamefont {Balz}},
  \bibinfo {author} {\bibfnamefont {A.~A.}\ \bibnamefont {Aczel}}, \bibinfo
  {author} {\bibfnamefont {B.}~\bibnamefont {Winn}}, \bibinfo {author}
  {\bibfnamefont {Y.}~\bibnamefont {Liu}}, \bibinfo {author} {\bibfnamefont
  {D.}~\bibnamefont {Pajerowski}}, \bibinfo {author} {\bibfnamefont
  {J.}~\bibnamefont {Yan}}, \bibinfo {author} {\bibfnamefont {C.~A.}\
  \bibnamefont {Bridges}}, \bibinfo {author} {\bibfnamefont {A.~T.}\
  \bibnamefont {Savici}}, \bibinfo {author} {\bibfnamefont {B.~C.}\
  \bibnamefont {Chakoumakos}}, \bibinfo {author} {\bibfnamefont {M.~D.}\
  \bibnamefont {Lumsden}}, \bibinfo {author} {\bibfnamefont {D.~A.}\
  \bibnamefont {Tennant}}, \bibinfo {author} {\bibfnamefont {R.}~\bibnamefont
  {Moessner}}, \bibinfo {author} {\bibfnamefont {D.~G.}\ \bibnamefont
  {Mandrus}},\ and\ \bibinfo {author} {\bibfnamefont {S.~E.}\ \bibnamefont
  {Nagler}},\ }\bibfield  {title} {\bibinfo {title} {Excitations in the
  field-induced quantum spin liquid state of $\alpha$-{RuCl}$_{3}$},\ }\href
  {https://doi.org/10.1038/s41535-018-0079-2} {\bibfield  {journal} {\bibinfo
  {journal} {npj. Quant. Mater.}\ }\textbf {\bibinfo {volume} {3}},\ \bibinfo
  {pages} {8} (\bibinfo {year} {2018})}\BibitemShut {NoStop}%
\bibitem [{\citenamefont {Kasahara}\ \emph {et~al.}(2018)\citenamefont
  {Kasahara}, \citenamefont {Ohnishi}, \citenamefont {Mizukami}, \citenamefont
  {Tanaka}, \citenamefont {Ma}, \citenamefont {Sugii}, \citenamefont {Kurita},
  \citenamefont {Tanaka}, \citenamefont {Nasu}, \citenamefont {Motome},
  \citenamefont {Shibauchi},\ and\ \citenamefont {Matsuda}}]{Kasahara2018}%
  \BibitemOpen
  \bibfield  {author} {\bibinfo {author} {\bibfnamefont {Y.}~\bibnamefont
  {Kasahara}}, \bibinfo {author} {\bibfnamefont {T.}~\bibnamefont {Ohnishi}},
  \bibinfo {author} {\bibfnamefont {Y.}~\bibnamefont {Mizukami}}, \bibinfo
  {author} {\bibfnamefont {O.}~\bibnamefont {Tanaka}}, \bibinfo {author}
  {\bibfnamefont {S.}~\bibnamefont {Ma}}, \bibinfo {author} {\bibfnamefont
  {K.}~\bibnamefont {Sugii}}, \bibinfo {author} {\bibfnamefont
  {N.}~\bibnamefont {Kurita}}, \bibinfo {author} {\bibfnamefont
  {H.}~\bibnamefont {Tanaka}}, \bibinfo {author} {\bibfnamefont
  {J.}~\bibnamefont {Nasu}}, \bibinfo {author} {\bibfnamefont {Y.}~\bibnamefont
  {Motome}}, \bibinfo {author} {\bibfnamefont {T.}~\bibnamefont {Shibauchi}},\
  and\ \bibinfo {author} {\bibfnamefont {Y.}~\bibnamefont {Matsuda}},\
  }\bibfield  {title} {\bibinfo {title} {Majorana quantization and half-integer
  thermal quantum $\rm{Hall}$ effect in a $\rm{Kitaev}$ spin liquid},\ }\href
  {https://doi.org/10.1038/s41586-018-0274-0} {\bibfield  {journal} {\bibinfo
  {journal} {Nature}\ }\textbf {\bibinfo {volume} {559}},\ \bibinfo {pages}
  {227} (\bibinfo {year} {2018})}\BibitemShut {NoStop}%
\bibitem [{\citenamefont {Yokoi}\ \emph {et~al.}(2021)\citenamefont {Yokoi},
  \citenamefont {Ma}, \citenamefont {Kasahara}, \citenamefont {Kasahara},
  \citenamefont {Shibauchi}, \citenamefont {Kurita}, \citenamefont {Tanaka},
  \citenamefont {Nasu}, \citenamefont {Motome}, \citenamefont {Hickey},
  \citenamefont {Trebst},\ and\ \citenamefont {Matsuda}}]{Yokoi2021}%
  \BibitemOpen
  \bibfield  {author} {\bibinfo {author} {\bibfnamefont {T.}~\bibnamefont
  {Yokoi}}, \bibinfo {author} {\bibfnamefont {S.}~\bibnamefont {Ma}}, \bibinfo
  {author} {\bibfnamefont {Y.}~\bibnamefont {Kasahara}}, \bibinfo {author}
  {\bibfnamefont {S.}~\bibnamefont {Kasahara}}, \bibinfo {author}
  {\bibfnamefont {T.}~\bibnamefont {Shibauchi}}, \bibinfo {author}
  {\bibfnamefont {N.}~\bibnamefont {Kurita}}, \bibinfo {author} {\bibfnamefont
  {H.}~\bibnamefont {Tanaka}}, \bibinfo {author} {\bibfnamefont
  {J.}~\bibnamefont {Nasu}}, \bibinfo {author} {\bibfnamefont {Y.}~\bibnamefont
  {Motome}}, \bibinfo {author} {\bibfnamefont {C.}~\bibnamefont {Hickey}},
  \bibinfo {author} {\bibfnamefont {S.}~\bibnamefont {Trebst}},\ and\ \bibinfo
  {author} {\bibfnamefont {Y.}~\bibnamefont {Matsuda}},\ }\bibfield  {title}
  {\bibinfo {title} {{Half-integer quantized anomalous thermal Hall effect in
  the Kitaev material candidate $\alpha$-RuCl$_3$}},\ }\href
  {https://doi.org/https://doi.org/10.1126/science.aay5551} {\bibfield
  {journal} {\bibinfo  {journal} {Science}\ }\textbf {\bibinfo {volume}
  {373}},\ \bibinfo {pages} {568} (\bibinfo {year} {2021})}\BibitemShut
  {NoStop}%
\bibitem [{\citenamefont {Yamashita}\ \emph {et~al.}(2020)\citenamefont
  {Yamashita}, \citenamefont {Gouchi}, \citenamefont {Uwatoko}, \citenamefont
  {Kurita},\ and\ \citenamefont {Tanaka}}]{Yamashita2020}%
  \BibitemOpen
  \bibfield  {author} {\bibinfo {author} {\bibfnamefont {M.}~\bibnamefont
  {Yamashita}}, \bibinfo {author} {\bibfnamefont {J.}~\bibnamefont {Gouchi}},
  \bibinfo {author} {\bibfnamefont {Y.}~\bibnamefont {Uwatoko}}, \bibinfo
  {author} {\bibfnamefont {N.}~\bibnamefont {Kurita}},\ and\ \bibinfo {author}
  {\bibfnamefont {H.}~\bibnamefont {Tanaka}},\ }\bibfield  {title} {\bibinfo
  {title} {{Sample dependence of half-integer quantized thermal Hall effect in
  the Kitaev spin-liquid candidate $\alpha$-RuCl$_3$}},\ }\href
  {https://doi.org/10.1103/PhysRevB.102.220404} {\bibfield  {journal} {\bibinfo
   {journal} {Phys. Rev. B}\ }\textbf {\bibinfo {volume} {102}},\ \bibinfo
  {pages} {220404} (\bibinfo {year} {2020})}\BibitemShut {NoStop}%
\bibitem [{\citenamefont {Bruin}\ \emph
  {et~al.}(2022{\natexlab{a}})\citenamefont {Bruin}, \citenamefont {Claus},
  \citenamefont {Matsumoto}, \citenamefont {Kurita}, \citenamefont {Tanaka},\
  and\ \citenamefont {Takagi}}]{Bruin2022}%
  \BibitemOpen
  \bibfield  {author} {\bibinfo {author} {\bibfnamefont {J.~A.~N.}\
  \bibnamefont {Bruin}}, \bibinfo {author} {\bibfnamefont {R.~R.}\ \bibnamefont
  {Claus}}, \bibinfo {author} {\bibfnamefont {Y.}~\bibnamefont {Matsumoto}},
  \bibinfo {author} {\bibfnamefont {N.}~\bibnamefont {Kurita}}, \bibinfo
  {author} {\bibfnamefont {H.}~\bibnamefont {Tanaka}},\ and\ \bibinfo {author}
  {\bibfnamefont {H.}~\bibnamefont {Takagi}},\ }\bibfield  {title} {\bibinfo
  {title} {{Robustness of the thermal Hall effect close to half-quantization in
  $\alpha$-RuCl$_3$}},\ }\href
  {https://doi.org/https://doi.org/10.1038/s41567-021-01501-y} {\bibfield
  {journal} {\bibinfo  {journal} {Nat. Phys.}\ }\textbf {\bibinfo {volume}
  {18}},\ \bibinfo {pages} {401} (\bibinfo {year}
  {2022}{\natexlab{a}})}\BibitemShut {NoStop}%
\bibitem [{\citenamefont {Kasahara}\ \emph {et~al.}(2022)\citenamefont
  {Kasahara}, \citenamefont {Suetsugu}, \citenamefont {Asaba}, \citenamefont
  {Kasahara}, \citenamefont {Shibauchi}, \citenamefont {Kurita}, \citenamefont
  {Tanaka},\ and\ \citenamefont {Matsuda}}]{Kasahara2022}%
  \BibitemOpen
  \bibfield  {author} {\bibinfo {author} {\bibfnamefont {Y.}~\bibnamefont
  {Kasahara}}, \bibinfo {author} {\bibfnamefont {S.}~\bibnamefont {Suetsugu}},
  \bibinfo {author} {\bibfnamefont {T.}~\bibnamefont {Asaba}}, \bibinfo
  {author} {\bibfnamefont {S.}~\bibnamefont {Kasahara}}, \bibinfo {author}
  {\bibfnamefont {T.}~\bibnamefont {Shibauchi}}, \bibinfo {author}
  {\bibfnamefont {N.}~\bibnamefont {Kurita}}, \bibinfo {author} {\bibfnamefont
  {H.}~\bibnamefont {Tanaka}},\ and\ \bibinfo {author} {\bibfnamefont
  {Y.}~\bibnamefont {Matsuda}},\ }\bibfield  {title} {\bibinfo {title}
  {{Quantized and unquantized thermal Hall conductance of the Kitaev spin
  liquid candidate $\alpha$-RuCl$_3$}},\ }\href
  {https://doi.org/10.1103/PhysRevB.106.L060410} {\bibfield  {journal}
  {\bibinfo  {journal} {Phys. Rev. B}\ }\textbf {\bibinfo {volume} {106}},\
  \bibinfo {pages} {L060410} (\bibinfo {year} {2022})}\BibitemShut {NoStop}%
\bibitem [{\citenamefont {Tanaka}\ \emph {et~al.}(2022)\citenamefont {Tanaka},
  \citenamefont {Mizukami}, \citenamefont {Harasawa}, \citenamefont
  {Hashimoto}, \citenamefont {Hwang}, \citenamefont {Kurita}, \citenamefont
  {Tanaka}, \citenamefont {Fujimoto}, \citenamefont {Matsuda}, \citenamefont
  {Moon},\ and\ \citenamefont {Shibauchi}}]{Tanaka2022}%
  \BibitemOpen
  \bibfield  {author} {\bibinfo {author} {\bibfnamefont {O.}~\bibnamefont
  {Tanaka}}, \bibinfo {author} {\bibfnamefont {Y.}~\bibnamefont {Mizukami}},
  \bibinfo {author} {\bibfnamefont {R.}~\bibnamefont {Harasawa}}, \bibinfo
  {author} {\bibfnamefont {K.}~\bibnamefont {Hashimoto}}, \bibinfo {author}
  {\bibfnamefont {K.}~\bibnamefont {Hwang}}, \bibinfo {author} {\bibfnamefont
  {N.}~\bibnamefont {Kurita}}, \bibinfo {author} {\bibfnamefont
  {H.}~\bibnamefont {Tanaka}}, \bibinfo {author} {\bibfnamefont
  {S.}~\bibnamefont {Fujimoto}}, \bibinfo {author} {\bibfnamefont
  {Y.}~\bibnamefont {Matsuda}}, \bibinfo {author} {\bibfnamefont {E.-G.}\
  \bibnamefont {Moon}},\ and\ \bibinfo {author} {\bibfnamefont
  {T.}~\bibnamefont {Shibauchi}},\ }\bibfield  {title} {\bibinfo {title}
  {{Thermodynamic evidence for a field-angle-dependent Majorana gap in a Kitaev
  spin liquid}},\ }\href
  {https://doi.org/https://doi.org/10.1038/s41567-021-01488-6} {\bibfield
  {journal} {\bibinfo  {journal} {Nat. Phys.}\ }\textbf {\bibinfo {volume}
  {18}},\ \bibinfo {pages} {429} (\bibinfo {year} {2022})}\BibitemShut
  {NoStop}%
\bibitem [{\citenamefont {Imamura}\ \emph
  {et~al.}(2024{\natexlab{a}})\citenamefont {Imamura}, \citenamefont
  {Suetsugu}, \citenamefont {Mizukami}, \citenamefont {Yoshida}, \citenamefont
  {Hashimoto}, \citenamefont {Ohtsuka}, \citenamefont {Kasahara}, \citenamefont
  {Kurita}, \citenamefont {Tanaka}, \citenamefont {Noh}, \citenamefont {Nasu},
  \citenamefont {Moon}, \citenamefont {Matsuda},\ and\ \citenamefont
  {Shibauchi}}]{Imamura2023}%
  \BibitemOpen
  \bibfield  {author} {\bibinfo {author} {\bibfnamefont {K.}~\bibnamefont
  {Imamura}}, \bibinfo {author} {\bibfnamefont {S.}~\bibnamefont {Suetsugu}},
  \bibinfo {author} {\bibfnamefont {Y.}~\bibnamefont {Mizukami}}, \bibinfo
  {author} {\bibfnamefont {Y.}~\bibnamefont {Yoshida}}, \bibinfo {author}
  {\bibfnamefont {K.}~\bibnamefont {Hashimoto}}, \bibinfo {author}
  {\bibfnamefont {K.}~\bibnamefont {Ohtsuka}}, \bibinfo {author} {\bibfnamefont
  {Y.}~\bibnamefont {Kasahara}}, \bibinfo {author} {\bibfnamefont
  {N.}~\bibnamefont {Kurita}}, \bibinfo {author} {\bibfnamefont
  {H.}~\bibnamefont {Tanaka}}, \bibinfo {author} {\bibfnamefont
  {P.}~\bibnamefont {Noh}}, \bibinfo {author} {\bibfnamefont {J.}~\bibnamefont
  {Nasu}}, \bibinfo {author} {\bibfnamefont {E.-G.}\ \bibnamefont {Moon}},
  \bibinfo {author} {\bibfnamefont {Y.}~\bibnamefont {Matsuda}},\ and\ \bibinfo
  {author} {\bibfnamefont {T.}~\bibnamefont {Shibauchi}},\ }\bibfield  {title}
  {\bibinfo {title} {{Majorana-fermion origin of the planar thermal Hall effect
  in the Kitaev magnet $\alpha$-RuCl$_3$}},\ }\href
  {https://doi.org/10.1126/sciadv.adk3539} {\bibfield  {journal} {\bibinfo
  {journal} {Science Advances}\ }\textbf {\bibinfo {volume} {10}},\ \bibinfo
  {pages} {eadk3539} (\bibinfo {year} {2024}{\natexlab{a}})}\BibitemShut
  {NoStop}%
\bibitem [{\citenamefont {Czajka}\ \emph {et~al.}(2023)\citenamefont {Czajka},
  \citenamefont {Gao}, \citenamefont {Hirschberger}, \citenamefont
  {Lampen-Kelley}, \citenamefont {Banerjee}, \citenamefont {Quirk},
  \citenamefont {Mandrus}, \citenamefont {Nagler},\ and\ \citenamefont
  {Ong}}]{czajka2023}%
  \BibitemOpen
  \bibfield  {author} {\bibinfo {author} {\bibfnamefont {P.}~\bibnamefont
  {Czajka}}, \bibinfo {author} {\bibfnamefont {T.}~\bibnamefont {Gao}},
  \bibinfo {author} {\bibfnamefont {M.}~\bibnamefont {Hirschberger}}, \bibinfo
  {author} {\bibfnamefont {P.}~\bibnamefont {Lampen-Kelley}}, \bibinfo {author}
  {\bibfnamefont {A.}~\bibnamefont {Banerjee}}, \bibinfo {author}
  {\bibfnamefont {N.}~\bibnamefont {Quirk}}, \bibinfo {author} {\bibfnamefont
  {D.~G.}\ \bibnamefont {Mandrus}}, \bibinfo {author} {\bibfnamefont {S.~E.}\
  \bibnamefont {Nagler}},\ and\ \bibinfo {author} {\bibfnamefont {N.~P.}\
  \bibnamefont {Ong}},\ }\bibfield  {title} {\bibinfo {title} {{Planar thermal
  Hall effect of topological bosons in the Kitaev magnet $\alpha$-RuCl$_3$}},\
  }\href {https://doi.org/https://doi.org/10.1038/s41563-022-01397-w}
  {\bibfield  {journal} {\bibinfo  {journal} {Nat. Mater.}\ }\textbf {\bibinfo
  {volume} {22}},\ \bibinfo {pages} {36} (\bibinfo {year} {2023})}\BibitemShut
  {NoStop}%
\bibitem [{\citenamefont {Lefran\ifmmode~\mbox{\c{c}}\else \c{c}\fi{}ois}\
  \emph {et~al.}(2022)\citenamefont {Lefran\ifmmode~\mbox{\c{c}}\else
  \c{c}\fi{}ois}, \citenamefont {Grissonnanche}, \citenamefont {Baglo},
  \citenamefont {Lampen-Kelley}, \citenamefont {Yan}, \citenamefont {Balz},
  \citenamefont {Mandrus}, \citenamefont {Nagler}, \citenamefont {Kim},
  \citenamefont {Kim}, \citenamefont {Doiron-Leyraud},\ and\ \citenamefont
  {Taillefer}}]{lefranccois2022}%
  \BibitemOpen
  \bibfield  {author} {\bibinfo {author} {\bibfnamefont {E.}~\bibnamefont
  {Lefran\ifmmode~\mbox{\c{c}}\else \c{c}\fi{}ois}}, \bibinfo {author}
  {\bibfnamefont {G.}~\bibnamefont {Grissonnanche}}, \bibinfo {author}
  {\bibfnamefont {J.}~\bibnamefont {Baglo}}, \bibinfo {author} {\bibfnamefont
  {P.}~\bibnamefont {Lampen-Kelley}}, \bibinfo {author} {\bibfnamefont {J.-Q.}\
  \bibnamefont {Yan}}, \bibinfo {author} {\bibfnamefont {C.}~\bibnamefont
  {Balz}}, \bibinfo {author} {\bibfnamefont {D.}~\bibnamefont {Mandrus}},
  \bibinfo {author} {\bibfnamefont {S.~E.}\ \bibnamefont {Nagler}}, \bibinfo
  {author} {\bibfnamefont {S.}~\bibnamefont {Kim}}, \bibinfo {author}
  {\bibfnamefont {Y.-J.}\ \bibnamefont {Kim}}, \bibinfo {author} {\bibfnamefont
  {N.}~\bibnamefont {Doiron-Leyraud}},\ and\ \bibinfo {author} {\bibfnamefont
  {L.}~\bibnamefont {Taillefer}},\ }\bibfield  {title} {\bibinfo {title}
  {{Evidence of a phonon Hall effect in the Kitaev spin liquid candidate
  $\alpha$-RuCl$_3$}},\ }\href {https://doi.org/10.1103/PhysRevX.12.021025}
  {\bibfield  {journal} {\bibinfo  {journal} {Phys. Rev. X}\ }\textbf {\bibinfo
  {volume} {12}},\ \bibinfo {pages} {021025} (\bibinfo {year}
  {2022})}\BibitemShut {NoStop}%
\bibitem [{\citenamefont {Czajka}\ \emph {et~al.}(2021)\citenamefont {Czajka},
  \citenamefont {Gao}, \citenamefont {Hirschberger}, \citenamefont
  {Lampen-Kelley}, \citenamefont {Banerjee}, \citenamefont {Yan}, \citenamefont
  {Mandrus}, \citenamefont {Nagler},\ and\ \citenamefont {Ong}}]{czajka2021}%
  \BibitemOpen
  \bibfield  {author} {\bibinfo {author} {\bibfnamefont {P.}~\bibnamefont
  {Czajka}}, \bibinfo {author} {\bibfnamefont {T.}~\bibnamefont {Gao}},
  \bibinfo {author} {\bibfnamefont {M.}~\bibnamefont {Hirschberger}}, \bibinfo
  {author} {\bibfnamefont {P.}~\bibnamefont {Lampen-Kelley}}, \bibinfo {author}
  {\bibfnamefont {A.}~\bibnamefont {Banerjee}}, \bibinfo {author}
  {\bibfnamefont {J.}~\bibnamefont {Yan}}, \bibinfo {author} {\bibfnamefont
  {D.~G.}\ \bibnamefont {Mandrus}}, \bibinfo {author} {\bibfnamefont {S.~E.}\
  \bibnamefont {Nagler}},\ and\ \bibinfo {author} {\bibfnamefont
  {N.}~\bibnamefont {Ong}},\ }\bibfield  {title} {\bibinfo {title}
  {{Oscillations of the thermal conductivity in the spin-liquid state of
  $\alpha$-RuCl$_3$}},\ }\href
  {https://doi.org/https://doi.org/10.1038/s41567-021-01243-x} {\bibfield
  {journal} {\bibinfo  {journal} {Nat. Phys.}\ }\textbf {\bibinfo {volume}
  {17}},\ \bibinfo {pages} {915} (\bibinfo {year} {2021})}\BibitemShut
  {NoStop}%
\bibitem [{\citenamefont {Zhang}\ \emph {et~al.}(2023)\citenamefont {Zhang},
  \citenamefont {Miao}, \citenamefont {Ward}, \citenamefont {Mandrus},
  \citenamefont {Nagler}, \citenamefont {McGuire},\ and\ \citenamefont
  {Yan}}]{Zhang2023}%
  \BibitemOpen
  \bibfield  {author} {\bibinfo {author} {\bibfnamefont {H.}~\bibnamefont
  {Zhang}}, \bibinfo {author} {\bibfnamefont {H.}~\bibnamefont {Miao}},
  \bibinfo {author} {\bibfnamefont {T.~Z.}\ \bibnamefont {Ward}}, \bibinfo
  {author} {\bibfnamefont {D.~G.}\ \bibnamefont {Mandrus}}, \bibinfo {author}
  {\bibfnamefont {S.~E.}\ \bibnamefont {Nagler}}, \bibinfo {author}
  {\bibfnamefont {M.~A.}\ \bibnamefont {McGuire}},\ and\ \bibinfo {author}
  {\bibfnamefont {J.}~\bibnamefont {Yan}},\ }\bibfield  {title} {\bibinfo
  {title} {{Anisotropy of thermal conductivity oscillations in relation to the
  Kitaev spin liquid phase}},\ }\href
  {https://doi.org/10.48550/arXiv.2310.03917} {\bibfield  {journal} {\bibinfo
  {journal} {arXiv:2310.03917}\ } (\bibinfo {year} {2023})}\BibitemShut
  {NoStop}%
\bibitem [{\citenamefont {Bruin}\ \emph
  {et~al.}(2022{\natexlab{b}})\citenamefont {Bruin}, \citenamefont {Claus},
  \citenamefont {Matsumoto}, \citenamefont {Nuss}, \citenamefont {Laha},
  \citenamefont {Lotsch}, \citenamefont {Kurita}, \citenamefont {Tanaka},\ and\
  \citenamefont {Takagi}}]{Bruuin_oscillation}%
  \BibitemOpen
  \bibfield  {author} {\bibinfo {author} {\bibfnamefont {J.~A.~N.}\
  \bibnamefont {Bruin}}, \bibinfo {author} {\bibfnamefont {R.~R.}\ \bibnamefont
  {Claus}}, \bibinfo {author} {\bibfnamefont {Y.}~\bibnamefont {Matsumoto}},
  \bibinfo {author} {\bibfnamefont {J.}~\bibnamefont {Nuss}}, \bibinfo {author}
  {\bibfnamefont {S.}~\bibnamefont {Laha}}, \bibinfo {author} {\bibfnamefont
  {B.~V.}\ \bibnamefont {Lotsch}}, \bibinfo {author} {\bibfnamefont
  {N.}~\bibnamefont {Kurita}}, \bibinfo {author} {\bibfnamefont
  {H.}~\bibnamefont {Tanaka}},\ and\ \bibinfo {author} {\bibfnamefont
  {H.}~\bibnamefont {Takagi}},\ }\bibfield  {title} {\bibinfo {title} {{Origin
  of oscillatory structures in the magnetothermal conductivity of the putative
  Kitaev magnet $\alpha$-RuCl$_3$}},\ }\href
  {https://doi.org/https://doi.org/10.1016/j.aop.2005.10.005} {\bibfield
  {journal} {\bibinfo  {journal} {APL Materials}\ }\textbf {\bibinfo {volume}
  {10}},\ \bibinfo {pages} {090703} (\bibinfo {year}
  {2022}{\natexlab{b}})}\BibitemShut {NoStop}%
\bibitem [{\citenamefont {Yan}\ and\ \citenamefont
  {McGuire}(2023)}]{Yan_SSVG_2023}%
  \BibitemOpen
  \bibfield  {author} {\bibinfo {author} {\bibfnamefont {J.-Q.}\ \bibnamefont
  {Yan}}\ and\ \bibinfo {author} {\bibfnamefont {M.~A.}\ \bibnamefont
  {McGuire}},\ }\bibfield  {title} {\bibinfo {title} {Self-selecting vapor
  growth of transition-metal-halide single crystals},\ }\href
  {https://doi.org/10.1103/PhysRevMaterials.7.013401} {\bibfield  {journal}
  {\bibinfo  {journal} {Phys. Rev. Mater.}\ }\textbf {\bibinfo {volume} {7}},\
  \bibinfo {pages} {013401} (\bibinfo {year} {2023})}\BibitemShut {NoStop}%
\bibitem [{\citenamefont {Zhang}\ \emph {et~al.}(2024)\citenamefont {Zhang},
  \citenamefont {McGuire}, \citenamefont {May}, \citenamefont {Chao},
  \citenamefont {Zheng}, \citenamefont {Chi}, \citenamefont {Sales},
  \citenamefont {Mandrus}, \citenamefont {Nagler}, \citenamefont {Miao},
  \citenamefont {Ye},\ and\ \citenamefont {Yan}}]{Zhang2024}%
  \BibitemOpen
  \bibfield  {author} {\bibinfo {author} {\bibfnamefont {H.}~\bibnamefont
  {Zhang}}, \bibinfo {author} {\bibfnamefont {M.~A.}\ \bibnamefont {McGuire}},
  \bibinfo {author} {\bibfnamefont {A.~F.}\ \bibnamefont {May}}, \bibinfo
  {author} {\bibfnamefont {H.-Y.}\ \bibnamefont {Chao}}, \bibinfo {author}
  {\bibfnamefont {Q.}~\bibnamefont {Zheng}}, \bibinfo {author} {\bibfnamefont
  {M.}~\bibnamefont {Chi}}, \bibinfo {author} {\bibfnamefont {B.~C.}\
  \bibnamefont {Sales}}, \bibinfo {author} {\bibfnamefont {D.~G.}\ \bibnamefont
  {Mandrus}}, \bibinfo {author} {\bibfnamefont {S.~E.}\ \bibnamefont {Nagler}},
  \bibinfo {author} {\bibfnamefont {H.}~\bibnamefont {Miao}}, \bibinfo {author}
  {\bibfnamefont {F.}~\bibnamefont {Ye}},\ and\ \bibinfo {author}
  {\bibfnamefont {J.}~\bibnamefont {Yan}},\ }\bibfield  {title} {\bibinfo
  {title} {{Stacking disorder and thermal transport properties of
  $\alpha$-RuCl$_3$}},\ }\href
  {https://doi.org/10.1103/PhysRevMaterials.8.014402} {\bibfield  {journal}
  {\bibinfo  {journal} {Phys. Rev. Mater.}\ }\textbf {\bibinfo {volume} {8}},\
  \bibinfo {pages} {014402} (\bibinfo {year} {2024})}\BibitemShut {NoStop}%
\bibitem [{\citenamefont {Wolf}\ \emph {et~al.}(2022)\citenamefont {Wolf},
  \citenamefont {Kaib}, \citenamefont {Razpopov}, \citenamefont {Biswas},
  \citenamefont {Riedl}, \citenamefont {Winter}, \citenamefont {Valent\'{\i}},
  \citenamefont {Saito}, \citenamefont {Hartmann}, \citenamefont {Vinokurova},
  \citenamefont {Doert}, \citenamefont {Isaeva}, \citenamefont {Bastien},
  \citenamefont {Wolter}, \citenamefont {B\"uchner},\ and\ \citenamefont
  {Lang}}]{Wolf2022}%
  \BibitemOpen
  \bibfield  {author} {\bibinfo {author} {\bibfnamefont {B.}~\bibnamefont
  {Wolf}}, \bibinfo {author} {\bibfnamefont {D.~A.~S.}\ \bibnamefont {Kaib}},
  \bibinfo {author} {\bibfnamefont {A.}~\bibnamefont {Razpopov}}, \bibinfo
  {author} {\bibfnamefont {S.}~\bibnamefont {Biswas}}, \bibinfo {author}
  {\bibfnamefont {K.}~\bibnamefont {Riedl}}, \bibinfo {author} {\bibfnamefont
  {S.~M.}\ \bibnamefont {Winter}}, \bibinfo {author} {\bibfnamefont
  {R.}~\bibnamefont {Valent\'{\i}}}, \bibinfo {author} {\bibfnamefont
  {Y.}~\bibnamefont {Saito}}, \bibinfo {author} {\bibfnamefont
  {S.}~\bibnamefont {Hartmann}}, \bibinfo {author} {\bibfnamefont
  {E.}~\bibnamefont {Vinokurova}}, \bibinfo {author} {\bibfnamefont
  {T.}~\bibnamefont {Doert}}, \bibinfo {author} {\bibfnamefont
  {A.}~\bibnamefont {Isaeva}}, \bibinfo {author} {\bibfnamefont
  {G.}~\bibnamefont {Bastien}}, \bibinfo {author} {\bibfnamefont {A.~U.~B.}\
  \bibnamefont {Wolter}}, \bibinfo {author} {\bibfnamefont {B.}~\bibnamefont
  {B\"uchner}},\ and\ \bibinfo {author} {\bibfnamefont {M.}~\bibnamefont
  {Lang}},\ }\bibfield  {title} {\bibinfo {title} {{Combined experimental and
  theoretical study of hydrostatic He-gas pressure effects in
  $\alpha$-RuCl$_3$}},\ }\href {https://doi.org/10.1103/PhysRevB.106.134432}
  {\bibfield  {journal} {\bibinfo  {journal} {Phys. Rev. B}\ }\textbf {\bibinfo
  {volume} {106}},\ \bibinfo {pages} {134432} (\bibinfo {year}
  {2022})}\BibitemShut {NoStop}%
\bibitem [{\citenamefont {Mi}\ \emph {et~al.}(2021)\citenamefont {Mi},
  \citenamefont {Wang}, \citenamefont {Gui}, \citenamefont {Pi}, \citenamefont
  {Zheng}, \citenamefont {Yang}, \citenamefont {Gan}, \citenamefont {Wang},
  \citenamefont {Li}, \citenamefont {Wang}, \citenamefont {Zhang},
  \citenamefont {Su}, \citenamefont {Chai},\ and\ \citenamefont {He}}]{Mi2015}%
  \BibitemOpen
  \bibfield  {author} {\bibinfo {author} {\bibfnamefont {X.}~\bibnamefont
  {Mi}}, \bibinfo {author} {\bibfnamefont {X.}~\bibnamefont {Wang}}, \bibinfo
  {author} {\bibfnamefont {H.}~\bibnamefont {Gui}}, \bibinfo {author}
  {\bibfnamefont {M.}~\bibnamefont {Pi}}, \bibinfo {author} {\bibfnamefont
  {T.}~\bibnamefont {Zheng}}, \bibinfo {author} {\bibfnamefont
  {K.}~\bibnamefont {Yang}}, \bibinfo {author} {\bibfnamefont {Y.}~\bibnamefont
  {Gan}}, \bibinfo {author} {\bibfnamefont {P.}~\bibnamefont {Wang}}, \bibinfo
  {author} {\bibfnamefont {A.}~\bibnamefont {Li}}, \bibinfo {author}
  {\bibfnamefont {A.}~\bibnamefont {Wang}}, \bibinfo {author} {\bibfnamefont
  {L.}~\bibnamefont {Zhang}}, \bibinfo {author} {\bibfnamefont
  {Y.}~\bibnamefont {Su}}, \bibinfo {author} {\bibfnamefont {Y.}~\bibnamefont
  {Chai}},\ and\ \bibinfo {author} {\bibfnamefont {M.}~\bibnamefont {He}},\
  }\bibfield  {title} {\bibinfo {title} {{Stacking faults in $\alpha$-RuCl$_3$
  revealed by local electric polarization}},\ }\href
  {https://doi.org/10.1103/PhysRevB.103.174413} {\bibfield  {journal} {\bibinfo
   {journal} {Phys. Rev. B}\ }\textbf {\bibinfo {volume} {103}},\ \bibinfo
  {pages} {174413} (\bibinfo {year} {2021})}\BibitemShut {NoStop}%
\bibitem [{\citenamefont {May}\ \emph {et~al.}(2020)\citenamefont {May},
  \citenamefont {Yan},\ and\ \citenamefont {McGuire}}]{May_2020}%
  \BibitemOpen
  \bibfield  {author} {\bibinfo {author} {\bibfnamefont {A.~F.}\ \bibnamefont
  {May}}, \bibinfo {author} {\bibfnamefont {J.}~\bibnamefont {Yan}},\ and\
  \bibinfo {author} {\bibfnamefont {M.~A.}\ \bibnamefont {McGuire}},\
  }\bibfield  {title} {\bibinfo {title} {{A practical guide for crystal growth
  of van der Waals layered materials}},\ }\href
  {https://doi.org/https://doi.org/10.1063/5.0015971} {\bibfield  {journal}
  {\bibinfo  {journal} {Journal of Applied Physics}\ }\textbf {\bibinfo
  {volume} {128}},\ \bibinfo {pages} {051101} (\bibinfo {year}
  {2020})}\BibitemShut {NoStop}%
\bibitem [{\citenamefont {Cao}\ \emph {et~al.}(2016)\citenamefont {Cao},
  \citenamefont {Banerjee}, \citenamefont {Yan}, \citenamefont {Bridges},
  \citenamefont {Lumsden}, \citenamefont {Mandrus}, \citenamefont {Tennant},
  \citenamefont {Chakoumakos},\ and\ \citenamefont {Nagler}}]{Cao2016}%
  \BibitemOpen
  \bibfield  {author} {\bibinfo {author} {\bibfnamefont {H.~B.}\ \bibnamefont
  {Cao}}, \bibinfo {author} {\bibfnamefont {A.}~\bibnamefont {Banerjee}},
  \bibinfo {author} {\bibfnamefont {J.-Q.}\ \bibnamefont {Yan}}, \bibinfo
  {author} {\bibfnamefont {C.~A.}\ \bibnamefont {Bridges}}, \bibinfo {author}
  {\bibfnamefont {M.~D.}\ \bibnamefont {Lumsden}}, \bibinfo {author}
  {\bibfnamefont {D.~G.}\ \bibnamefont {Mandrus}}, \bibinfo {author}
  {\bibfnamefont {D.~A.}\ \bibnamefont {Tennant}}, \bibinfo {author}
  {\bibfnamefont {B.~C.}\ \bibnamefont {Chakoumakos}},\ and\ \bibinfo {author}
  {\bibfnamefont {S.~E.}\ \bibnamefont {Nagler}},\ }\bibfield  {title}
  {\bibinfo {title} {{Low-temperature crystal and magnetic structure of
  $\alpha$-RuCl$_3$}},\ }\href {https://doi.org/10.1103/PhysRevB.93.134423}
  {\bibfield  {journal} {\bibinfo  {journal} {Phys. Rev. B}\ }\textbf {\bibinfo
  {volume} {93}},\ \bibinfo {pages} {134423} (\bibinfo {year}
  {2016})}\BibitemShut {NoStop}%
\bibitem [{\citenamefont {Kim}\ \emph {et~al.}(2022)\citenamefont {Kim},
  \citenamefont {Yuan},\ and\ \citenamefont {Kim}}]{Kim2022}%
  \BibitemOpen
  \bibfield  {author} {\bibinfo {author} {\bibfnamefont {S.}~\bibnamefont
  {Kim}}, \bibinfo {author} {\bibfnamefont {B.}~\bibnamefont {Yuan}},\ and\
  \bibinfo {author} {\bibfnamefont {Y.-J.}\ \bibnamefont {Kim}},\ }\bibfield
  {title} {\bibinfo {title} {{$\alpha$-RuCl$_3$ and other Kitaev materials}},\
  }\href {https://doi.org/10.1063/5.0101512} {\bibfield  {journal} {\bibinfo
  {journal} {APL Materials}\ }\textbf {\bibinfo {volume} {10}},\ \bibinfo
  {pages} {080903} (\bibinfo {year} {2022})}\BibitemShut {NoStop}%
\bibitem [{\citenamefont {Pott}\ and\ \citenamefont
  {Schefzyk}(1983)}]{pott1983apparatus}%
  \BibitemOpen
  \bibfield  {author} {\bibinfo {author} {\bibfnamefont {R.}~\bibnamefont
  {Pott}}\ and\ \bibinfo {author} {\bibfnamefont {R.}~\bibnamefont
  {Schefzyk}},\ }\bibfield  {title} {\bibinfo {title} {{Apparatus for measuring
  the thermal expansion of solids between 1.5 and 380\,K}},\ }\href
  {https://doi.org/10.1088/0022-3735/16/5/018} {\bibfield  {journal} {\bibinfo
  {journal} {Journal of Physics E: Scientific Instruments}\ }\textbf {\bibinfo
  {volume} {16}},\ \bibinfo {pages} {444} (\bibinfo {year} {1983})}\BibitemShut
  {NoStop}%
\bibitem [{\citenamefont {Nagai}\ \emph {et~al.}(2020)\citenamefont {Nagai},
  \citenamefont {Jinno}, \citenamefont {Yoshitake}, \citenamefont {Nasu},
  \citenamefont {Motome}, \citenamefont {Itoh},\ and\ \citenamefont
  {Shimizu}}]{Nagai_NQRNMR_2020}%
  \BibitemOpen
  \bibfield  {author} {\bibinfo {author} {\bibfnamefont {Y.}~\bibnamefont
  {Nagai}}, \bibinfo {author} {\bibfnamefont {T.}~\bibnamefont {Jinno}},
  \bibinfo {author} {\bibfnamefont {J.}~\bibnamefont {Yoshitake}}, \bibinfo
  {author} {\bibfnamefont {J.}~\bibnamefont {Nasu}}, \bibinfo {author}
  {\bibfnamefont {Y.}~\bibnamefont {Motome}}, \bibinfo {author} {\bibfnamefont
  {M.}~\bibnamefont {Itoh}},\ and\ \bibinfo {author} {\bibfnamefont
  {Y.}~\bibnamefont {Shimizu}},\ }\bibfield  {title} {\bibinfo {title}
  {{Two-step gap opening across the quantum critical point in the Kitaev
  honeycomb magnet $\alpha$-RuCl$_3$}},\ }\href
  {https://doi.org/10.1103/PhysRevB.101.020414} {\bibfield  {journal} {\bibinfo
   {journal} {Phys. Rev. B}\ }\textbf {\bibinfo {volume} {101}},\ \bibinfo
  {pages} {020414} (\bibinfo {year} {2020})}\BibitemShut {NoStop}%
\bibitem [{\citenamefont {Kim}\ \emph {et~al.}(2024)\citenamefont {Kim},
  \citenamefont {Horsley}, \citenamefont {Ruff}, \citenamefont {Moreno},\ and\
  \citenamefont {Kim}}]{Kim2024}%
  \BibitemOpen
  \bibfield  {author} {\bibinfo {author} {\bibfnamefont {S.}~\bibnamefont
  {Kim}}, \bibinfo {author} {\bibfnamefont {E.}~\bibnamefont {Horsley}},
  \bibinfo {author} {\bibfnamefont {J.~P.~C.}\ \bibnamefont {Ruff}}, \bibinfo
  {author} {\bibfnamefont {B.~D.}\ \bibnamefont {Moreno}},\ and\ \bibinfo
  {author} {\bibfnamefont {Y.-J.}\ \bibnamefont {Kim}},\ }\bibfield  {title}
  {\bibinfo {title} {{Structural transition and magnetic anisotropy in
  $\alpha$-RuCl$_3$}},\ }\href {https://doi.org/10.1103/PhysRevB.109.L140101}
  {\bibfield  {journal} {\bibinfo  {journal} {Phys. Rev. B}\ }\textbf {\bibinfo
  {volume} {109}},\ \bibinfo {pages} {L140101} (\bibinfo {year}
  {2024})}\BibitemShut {NoStop}%
\bibitem [{\citenamefont {Kurumaji}(2023)}]{Kurumaji2023}%
  \BibitemOpen
  \bibfield  {author} {\bibinfo {author} {\bibfnamefont {T.}~\bibnamefont
  {Kurumaji}},\ }\bibfield  {title} {\bibinfo {title} {{Symmetry-based
  requirement for the measurement of electrical and thermal Hall conductivity
  under an in-plane magnetic field}},\ }\href
  {https://doi.org/10.1103/PhysRevResearch.5.023138} {\bibfield  {journal}
  {\bibinfo  {journal} {Phys. Rev. Res.}\ }\textbf {\bibinfo {volume} {5}},\
  \bibinfo {pages} {023138} (\bibinfo {year} {2023})}\BibitemShut {NoStop}%
\bibitem [{\citenamefont {He}\ \emph {et~al.}(2018)\citenamefont {He},
  \citenamefont {Wang}, \citenamefont {Wang}, \citenamefont {Hardy},
  \citenamefont {Wolf}, \citenamefont {Adelmann}, \citenamefont {Brckel},
  \citenamefont {Su},\ and\ \citenamefont {Meingast}}]{He_2018}%
  \BibitemOpen
  \bibfield  {author} {\bibinfo {author} {\bibfnamefont {M.}~\bibnamefont
  {He}}, \bibinfo {author} {\bibfnamefont {X.}~\bibnamefont {Wang}}, \bibinfo
  {author} {\bibfnamefont {L.}~\bibnamefont {Wang}}, \bibinfo {author}
  {\bibfnamefont {F.}~\bibnamefont {Hardy}}, \bibinfo {author} {\bibfnamefont
  {T.}~\bibnamefont {Wolf}}, \bibinfo {author} {\bibfnamefont {P.}~\bibnamefont
  {Adelmann}}, \bibinfo {author} {\bibfnamefont {T.}~\bibnamefont {Brckel}},
  \bibinfo {author} {\bibfnamefont {Y.}~\bibnamefont {Su}},\ and\ \bibinfo
  {author} {\bibfnamefont {C.}~\bibnamefont {Meingast}},\ }\bibfield  {title}
  {\bibinfo {title} {{Uniaxial and hydrostatic pressure effects in
  $\alpha$-RuCl$_3$ single crystals via thermal-expansion measurements}},\
  }\href {https://doi.org/10.1088/1361-648X/aada1e} {\bibfield  {journal}
  {\bibinfo  {journal} {Journal of Physics: Condensed Matter}\ }\textbf
  {\bibinfo {volume} {30}},\ \bibinfo {pages} {385702} (\bibinfo {year}
  {2018})}\BibitemShut {NoStop}%
\bibitem [{\citenamefont {Morosin}\ and\ \citenamefont
  {Narath}(2004)}]{Morosin_1964}%
  \BibitemOpen
  \bibfield  {author} {\bibinfo {author} {\bibfnamefont {B.}~\bibnamefont
  {Morosin}}\ and\ \bibinfo {author} {\bibfnamefont {A.}~\bibnamefont
  {Narath}},\ }\bibfield  {title} {\bibinfo {title} {X-ray diffraction and
  nuclear quadrupole resonance studies of chromium trichloride},\ }\href
  {https://doi.org/10.1063/1.1725428} {\bibfield  {journal} {\bibinfo
  {journal} {The Journal of Chemical Physics}\ }\textbf {\bibinfo {volume}
  {40}},\ \bibinfo {pages} {1958} (\bibinfo {year} {2004})}\BibitemShut
  {NoStop}%
\bibitem [{\citenamefont {Imamura}\ \emph
  {et~al.}(2024{\natexlab{b}})\citenamefont {Imamura}, \citenamefont
  {Mizukami}, \citenamefont {Tanaka}, \citenamefont {Grasset}, \citenamefont
  {Konczykowski}, \citenamefont {Kurita}, \citenamefont {Tanaka}, \citenamefont
  {Matsuda}, \citenamefont {Yamada}, \citenamefont {Hashimoto},\ and\
  \citenamefont {Shibauchi}}]{imamura2023irradiation}%
  \BibitemOpen
  \bibfield  {author} {\bibinfo {author} {\bibfnamefont {K.}~\bibnamefont
  {Imamura}}, \bibinfo {author} {\bibfnamefont {Y.}~\bibnamefont {Mizukami}},
  \bibinfo {author} {\bibfnamefont {O.}~\bibnamefont {Tanaka}}, \bibinfo
  {author} {\bibfnamefont {R.}~\bibnamefont {Grasset}}, \bibinfo {author}
  {\bibfnamefont {M.}~\bibnamefont {Konczykowski}}, \bibinfo {author}
  {\bibfnamefont {N.}~\bibnamefont {Kurita}}, \bibinfo {author} {\bibfnamefont
  {H.}~\bibnamefont {Tanaka}}, \bibinfo {author} {\bibfnamefont
  {Y.}~\bibnamefont {Matsuda}}, \bibinfo {author} {\bibfnamefont {M.~G.}\
  \bibnamefont {Yamada}}, \bibinfo {author} {\bibfnamefont {K.}~\bibnamefont
  {Hashimoto}},\ and\ \bibinfo {author} {\bibfnamefont {T.}~\bibnamefont
  {Shibauchi}},\ }\bibfield  {title} {\bibinfo {title} {{Defect-induced
  low-energy Majorana excitations in the Kitaev magnet $\alpha$-RuCl$_3$}},\
  }\href {https://doi.org/10.1103/PhysRevX.14.011045} {\bibfield  {journal}
  {\bibinfo  {journal} {Phys. Rev. X}\ }\textbf {\bibinfo {volume} {14}},\
  \bibinfo {pages} {011045} (\bibinfo {year} {2024}{\natexlab{b}})}\BibitemShut
  {NoStop}%
\end{thebibliography}%

\end{document}